\documentclass[twocolumn]{autart}    
\usepackage{subfigure}
\usepackage{amssymb}
\usepackage{amsfonts}
\usepackage{amsmath}
\usepackage{graphicx}
\usepackage{color}
\usepackage{mathrsfs}
\usepackage{graphicx}          

\begin{document}

\begin{frontmatter}

\title{Secure distributed adaptive optimal coordination of nonlinear cyber-physical systems with attack diagnosis\thanksref{footnoteinfo}} 

\thanks[footnoteinfo]{This paper was not presented at any IFAC
meeting. Corresponding author: Guang-Hong~Yang. Tel. +XXXIX-VI-mmmxxi.
Fax +XXXIX-VI-mmmxxv.}

\author[Paestum]{Liwei An}\ead{liwei.an@foxmail.com},    
\author[Paestum,Rome]{Guang-Hong Yang}\ead{yangguanghong@ise.neu.edu.cn}               

\address[Paestum]{College of Information Science and Engineering, Northeastern University, Shenyang 110819, P.R.China}  
\address[Rome]{State Key Laboratory of Synthetical Automation
for Process Industries, Northeastern University, Shenyang 110819, P.R.China}


\begin{abstract}                          
This paper studies the problem of distributed optimal coordination (DOC) for a class of nonlinear large-scale cyber-physical systems (CPSs) in the presence of cyber attacks. A secure DOC architecture with attack diagnosis is proposed that guarantees the attack-free subsystems to achieve the output consensus which minimizes the sum of their objective functions, while the attacked subsystems converge to preset secure states. A two-layer DOC structure is established with emphasis on the interactions between cyber and physical layers, where a command-driven control law is designed that generates provable optimal output consensus. Differing from the existing fault diagnosis methods which are generally applicable to given failure types, the focus of the attack diagnosis is to achieve detection and isolation for {\it arbitrary} malicious behaviors. To this end, double coupling residuals are generated by a carefully-designed distributed filter. The adaptive thresholds with prescribed performance are designed to enhance the detectability and isolability. It is theoretically guaranteed that any attack signal cannot bypass the designed attack diagnosis methodology to destroy the convergence of the DOC algorithm, and the locally-occurring detectable attack can be isolated from the propagating attacks from neighboring subsystems. Simulation results for the motion coordination of multiple remotely operated underwater vehicles illustrate the effectiveness of the proposed architecture.
\end{abstract}

\end{frontmatter}

\section{Introduction}

Recently, Cyber-Physical Systems (CPSs), which closely connect cyber and physical worlds, have gained much research interest in many fields, such as computer field, control field, battle field. By utilizing the close interaction between cyber and physical parts, the ``intelligence'' of physical systems can be sufficiently enhanced in order to fulfil some complex, precise or dangerous tasks, such as remote diagnosis, deep sea exploration \cite{PA2014}. However, the networked connection between cyber and physical parts also often leads to large attack space, such that the CPSs are vulnerable to various types of adversarial attacks. Some famous examples such as the Maroochy water breach \cite{JS2007} and Stuxnet \cite{JP2010} indicate the CPS security as a fundamental issue to be studied.


With potential applications of distributed optimization in large-scale CPSs \cite{SS2018}, many important results on discrete- or continuous-time DO algorithms have been reported \cite{AN2009,IL2011,MZ2012,BJ2008,BG2014}. In these algorithms, each individual (or said agent in multi-agent systems) only performs the designed optimization dynamics, ignoring its own dynamics. Note that the physical dynamic systems are usually indispensable parts for achieving DO task, such as the cooperative search of radio sources \cite{CY2014}, the motion coordination \cite{YX2019} and the distributed optimal power flow \cite{ED2013,SB2015}. Hence, it is relevant to study the distributed optimization problems together with physical dynamics, termed {\it distributed optimal coordination (DOC)}. In fact, the DOC can be completed based on the CPS architecture by effectively combining of cyber computation/communication and physical dynamics/control \cite{YZZ2017}. Recently, many important results for DOC have been reported for multi-agent system with various physical dynamics by designing integrated closed-loop control laws, such as integrator-type dynamics \cite{PL2014,YX2019,YZ2015}, continuous-time linear dynamics \cite{YZ2017,ZL2019}, Euler-Lagrangian dynamics \cite{YZZ2017}. More references for DOC can be found in \cite{TY2019}. Motivated by these and considering the non-ignorable nonlinear uncertain dynamics in many physical agents \cite{WW2017,WL2017,YZZ2017}, this paper investigates the DOC problem for a class of certain nonlinear large-scale systems on the CPS platforms.

Given the growing threat of malicious attacks in large-scale (and safety-critical) CPSs, the vulnerability of consensus-based DOC algorithms is also with respect to cyber attacks. Hence, the other main objective of this paper is to address the issue of security of consensus-based DOC dynamics by providing certain safety guarantees based on attack diagnosis. The recent works \cite{LS2015,SS2018,CZ2019} also consider the problem of resilient DO under different adversarial models than the ones that we consider here, and the agent's own physical dynamics is not considered there. Other related important works on distributed/decentralized sensor fault diagnosis and secure state estimation against sensor attacks for large-scale CPSs have been reported in \cite{QZ2012,VR2015-2,VR2015,LZ2019,FP2013,LW2019,JZ2015}. In the existing fault diagnosis and fault-tolerant results, the fault detectors are in general designed for given failure types \cite{MM1989,HF2014}, such as loss of effectiveness \cite{LZ2019}, bias faults \cite{QZ2012,VR2015-2,VR2015}. As we will see later, in our problem formulation the attack model can be considered to contain infinite number of failure types and one cannot afford to construct a fault detector for each possible failure type. In \cite{FP2013,LW2019,JZ2015}, the attack-resilient mechanisms are designed in the presence of arbitrary adversarial behaviors under the framework of distributed estimation, outside of DOC framework.

In this paper, we propose a secure DOC architecture for a class of nonlinear large-scale CPSs in the presence of cyber attacks. The overall architecture consists of cyber and physical parts, and each physical subsystem is modeled as a nonlinear parametric strict-feedback system equipped with a dedicated decision-making agent in the cyber superstratum. The cyber core (multi-agent network) focuses on the design of DOC and attack diagnosis, and the physical part performs the corresponding optimization task following the cyber-core control command. The objective is to steer the physical systems to achieve the output consensus at the minimizer of a given team performance function in a distributed fashion and provide certain safety guarantees based on attack diagnosis. The contributions of this paper are threefold.

The first contribution of this paper is, differing from the existing integrated closed-loop control schemes proposed in \cite{YZZ2017,PL2014,YX2019,YZ2015,YZ2017,ZL2019}, to propose a two-layer DOC structure where the cyber-layer optimizer generates a control command which is transmitted to the physical-layer control for local regulation. To overcome the difficulty caused by the dynamic mismatch between the traditional DO algorithms \cite{AN2009,IL2011,MZ2012,BJ2008,BG2014} and adaptive backstepping control systems \cite{MK1995,HO2017,XD2003,WW2010}, a novel command-driven control strategy is designed. It is proved that the proposed algorithm ensures all subsystems to achieve the optimal consensus under the healthy (attack-free) environment.

As the second contribution, we provide the design and analysis of an attack detection and isolation (ADI) methodology. The existing fault diagnosis schemes are usually designed for given fault types \cite{MM1989,HF2014} and cannot guarantee the detectability for {\it arbitrary} malicious behaviors in theory. Also, due to the coupling effects of multiple propagated attacks on the physical dynamics and cyber dynamics which are interacted, the attack isolation becomes more challenging. To this end, double coupling residuals and analytical redundancy relations (ARRs) are generated by a carefully-designed distributed filter. The adaptive thresholds with prescribed performance are designed
to enhance the detectability and isolability. It is theoretically guaranteed that {\it any} attack signal cannot bypass the ADI methodology to destroy the convergence of the DOC algorithm, and the locally-occurring detectable attacks can be isolated from the propagating attacks from the neighboring subsystems.

The last contribution is to develop a secure version of the DOC protocol based on the ADI methodology, which can provide a safety guarantee in the sense that the healthy physical subsystems (satisfying ARRs) reach the output consensus at the optimal solution of the sum of their objective functions, while the attacked physical subsystems (not
satisfying ARRs) converge to preset secure states.


\section{Preliminaries}

\subsection{Notations}

The symbols $\mathbb{R}$ and $\mathbb{B}$ denote the set of real and Boolean numbers, respectively. $\mathbb{C}^{n}_m$ represents the set of $n$-order differentiable $m$-dimension function vectors. The cardinality of a set $\mathbb{S}$ is denoted by $|\mathbb{S}|$. $\otimes$ and $\circ$ stand for the Kronecker product and Hadamard product, respectively. For a given time interval $\Xi$, $\nu(\Xi)$ represents its Lebesgue measure. $\mathrm{sgn}(\cdot)$ represents the sign function. Denote $1_N=[1,\cdots,1]^T\in \mathbb{R}^N$. For a vector sequence $\{Y^{(j)}\}_{j=1}^N$, we denote the notation $Y=\mathrm{vec}(Y^{(1)},\cdots,$ $Y^{(N)})$ if not specified.  

\subsection{Graph theory}

A weighted undirected graph $\mathcal{G}=(\mathcal{V},\mathcal{E},A)$ consists of $N$ vertices (or nodes, or agents in multi-agent networks) $\mathcal{V}=\{v_1,\cdots,v_N\}$, a set of edges (or links) $\mathcal{E}\subset \mathcal{V}\times \mathcal{V}$ and an adjacency matrix $A=\{w_{ij}\}_{N\times N}$ with nonnegative element $w_{ij}>0$ if $(v_i,v_j)\in\mathcal{E}$ and $w_{ij}=0$ otherwise. The neighbors of vertex $v_i\in \mathcal{V}$ are denoted by the set $\mathbf{N}_i=\{v_j\in \mathcal{V}:(v_j,v_i)\in \mathcal{E}\}$. The Laplacian matrix $\mathcal{L} =(l_{ij})_{N\times N}$ associated with
 graph $\mathcal{G}$ is defined as $l_{ii}=\sum_{j=1}^Nw_{ij}$ and $l_{ij}=-w_{ij}$ for $i\ne j$. For an undirected graph, the matrix $\mathcal{L}$ is symmetric and semi-positive. A path from vertex $v_i$ to vertex $v_j$ in graph $\mathcal{G}$ is a sequence of edges
$(v_i, v_{i_1}), (v_{i_1} ,v_{i_2}),\cdots, (v_{i_k} , v_j )$ in the graph with distinct nodes
$v_{i_k}\in \mathcal{V}$. An undirected graph is connected if there is a path from every vertex to other vertex in the graph.

\section{DOC architecture}

Consider a CPS consisting of $N$ subsystems, which aims at achieving the DOC task. The $j$th subsystem, $j=1,\cdots,N$ is described by the pair $(\mathcal{P}^{(j)},\mathcal{C}^{(j)})$, where $\mathcal{C}^{(j)}$ denotes the cyber part which is responsible for task decision-making, while $\mathcal{P}^{(j)}$ denotes the physical part which is responsible for task execution. The physical part $\mathcal{P}^{(j)}$ is modeled as a nonlinear dynamical system
\begin{equation}
\Sigma^{(j)}:\left\{\begin{aligned}
\dot{x}_i^{(j)}(t)=&x_{i+1}^{(j)}(t)+\varphi_i^{(j)}(\bar x_i^{(j)}(t))\theta_j\\
\dot{x}_n^{(j)}(t)=&\beta_ju^{(j)}(t)+\varphi_n^{(j)}(x^{(j)}(t))\theta_j,\\
y^{(j)}(t)=&x_1^{(j)}(t)+a^{(j)}(t)
\end{aligned}\right.
\end{equation}
where $i=1,\cdots,n-1$, $x_i^{(j)}(t)\in \mathbb{R}^m$, $\bar x_i^{(j)}(t)=\mathrm{vec}(x_1^{(j)}(t),\cdots,x^{(j)}_i(t))$ $\in \mathbb{R}^{im}$, $x^{(j)}(t)=\mathrm{vec}(x^{(j)}_1(t),\cdots,$ $x^{(j)}_n(t))\in \mathbb{R}^{nm}$ is the measurable state; $u^{(j)}(t)\in \mathbb{R}^m$ is the control input; $\varphi_i^{(j)}(\bar x^{(j)}(t))\in \mathbb{R}^{m\times p}$ and $\beta_j\in \mathbb{R}^{m\times m}$ are known nonsingular matrix; $\theta_j\in \mathbb{R}^p$ is unknown constant; $y^{(j)}(t)\in \mathbb{R}^m$ is the output measurement transmitted to the cyber superstratum through a wireless network channel and $a^{(j)}(t)\in \mathbb{R}^m$ denotes the cyber attacks corrupting the sensor transmitting signal. In particular, in order to provide security guarantees against {\it worst case} adversarial behavior, we allow the adversarial attacker to know the overall system model, system state, control input and the possible fault detector $\mathfrak{D}$ (e.g., distributed adaptive observers \cite{QZ2012,VR2015-2,VR2015}) equipped on the CPS. Thus, the attack signal can be modeled as
$$
a^{(j)}(t)=\kappa^{(j)}(t-T_a^{(j)})\phi^{(j)}(x(t),u(t),\mathfrak{D},t-T_a^{(j)})
$$
where $\kappa^{(j)}(t)$ is the time profile and $\phi^{(j)}(\cdot,\cdot,\cdot,\cdot)\in \mathbb{R}^m$ is an unknown function that occurs at the unknown time instant $T_a^{(j)}$. We make no assumption on $\phi^{(j)}(\cdot,\cdot,\cdot,\cdot)$, which may be any (such as unbounded, discontinuous) function vector. The time profile of the attack is modeled as $\kappa^{(j)}(r)=0$ if $r<0$ and $\kappa^{(j)}(r)=1$ otherwise. Multiple cyber attacks may occur simultaneously or sequentially, for example, $T_a^{(1)}\le\cdots\le T_a^{(j')}$ with $j'\le N$.

{\bf Remark 1.} System (1) can represent many practical systems such as mobile robots, chemical reactors, wind tunnels, and autonomous vehicles \cite{MK1995}. Some extensive researches for system (1) in presence of external disturbances, actuator failures, etc., have been studied well \cite{HO2017,XD2003,WW2010} (these are easily extended to the current framework and thus no longer considered here). Particularly, the works in \cite{WW2017,WL2017} investigated the adaptive  leader-following consensus control of system (1). In this paper, the problem of DOC further minimizes a given team performance function on the basis of consensus.

{\bf Remark 2.} In general, the fault detector $\mathfrak{D}$ is designed for given failure types and cannot guarantee the detectability for {\it arbitrary} malicious behaviors in theory. Due to adversary's strategic design, here we can assume that the attack signal $a^{(j)}(t)$ in system (1) denotes a {\it strategic} attack model which can potentially bypass the fault detector $\mathfrak{D}$ to destroy the system convergence based on the knowledge of system and detector $\mathfrak{D}$ (see stealthy attack design methods against various fault detectors, e.g., \cite{LW2018,YL2009,YC2019,TY2020} and references therein).

\begin{figure}
  \centering
  \includegraphics[width=7cm]{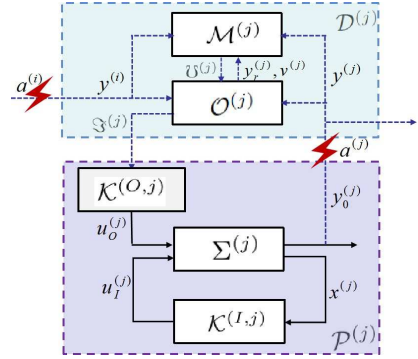}\\
  \caption{Secure DOC architecture of the $j$th subsystem of CPS affected by cyber attacks.}
\end{figure}

The overall DOC architecture of the considered CPS is illustrated in Fig. 1. Similar CPS architectures can also be found in \cite{MZ2015,VR2015}. The cyber part $\mathcal{C}^{(j)}$ consists of a decision-making multi-agent network. Each decision-making agent, denoted by $\mathcal{D}^{(j)}$, is responsible for sending the control command $\Im^{(j)}$ to $\Sigma^{(j)}$. The agent $\mathcal{D}^{(j)}$ contains an optimization module and a monitoring module, denoted by $\mathcal{O}^{(j)}$ and $\mathcal{M}^{(j)}$, respectively. The module $\mathcal{O}^{(j)}$ is used to optimize its local objective function, while exchanging its information with its neighbors under a network topology $\mathcal{G}$. Since the adversarial attackers can perform the attack $a^{(j)}(t)$ to corrupt the communication $y^{(j)}(t)$ from physical stratum to cyber stratum, a cyber attack in the subsystem $(\mathcal{P}^{(j)},\mathcal{C}^{(j)})$ can also be propagated to neighbors via the information exchange between the agent $\mathcal{D}^{(j)}$ and its neighboring agents. This  complicates the identification of attacked subsystems. To address it, each module $\mathcal{M}^{(j)}$ is used to detect and isolate the cyber attack $a^{(j)}$.
In the physical part $\mathcal{P}^{(j)}$, the control module $\mathcal{K}^{(j)}$ consisting of an inner-loop stabilizing module $\mathcal{K}^{(I,j)}$ and an outer-loop tracking module $\mathcal{K}^{(O,j)}$ drives the physical dynamics in accordance with the control command $\Im^{(j)}$ coming from the decision-making agent $\mathcal{D}^{(j)}$. Fig. 1 illustrates that {\it sufficient interaction} between cyber and physical parts in the CPS architecture.

{\bf Objective of this paper:} Design the decision-making agent $\mathcal{D}^{(j)}$ (including the optimization module $\mathcal{O}^{(j)}$ and the monitoring module $\mathcal{M}^{(j)}$) and the control agent $\mathcal{K}^{(j)}$, such that

1) {\bf Optimality:} Under the healthy (attack-free) condition, all the physical subsystems cooperatively reach the optimal output that minimizes the following team performance function:
\begin{equation}
\min\sum_{j=1}^Ng^{(j)}(s),~s\in \mathbb{R}^m
\end{equation}
where $g^{(j)}: \mathbb{R}^m\to \mathbb{R}$ is a local performance function privately known to the agent $\mathcal{D}^{(j)}$; and all the closed-loop signals of the physical system are uniformly ultimately bounded (UUB).

2) {\bf Security:} Detect and isolate multiple cyber attacks occurring at the network communication between physical stratum and cyber superstratum, and guarantee that the attacked subsystem $\mathcal{P}^{(j)}$, $j\in \mathbf{N}_a$ converges to a given secure state $y_s^{(j)}$, while the healthy subsystems achieve the output consensus at the optimal solution of
\begin{equation}
\min\sum_{j\in \mathbf{N}_h}g^{(j)}(s),~s\in \mathbb{R}^m
\end{equation}
where $\mathbf{N}_h$ represents the index set of the healthy subsystems, i.e., $\mathbf{N}_h=\{j\in\{1,\cdots,N\}:a^{(j)}(t)=0,\forall t\ge 0\}$, and $\mathbf{N}_a=\{1,\cdots,N\}\backslash\mathbf{N}_h$.

Here we consider the scenario that the communication from physical stratum to cyber superstratum can be attacked, while the communication from the cyber superstratum to the physical stratum is secure \cite{TY2020}. For example, for the GPS spoofing attacks on multiple Unmanned Aerial Vehicles (UAVs) \cite{AE2020}, the GPS attacker in fact can tamper with the location information transmitted from the UAV to the ground station, but cannot tamper with the control command from the ground station to the UAV or other communications among the ground stations.
Indeed, in many practical situations, the adversarial attackers may also tamper with the control command $\Im^{(j)}$ transmitted from cyber superstratum to physical stratum or communication among decision-making agents in the cyber superstratum; however, the design of the attack diagnosis and attack-tolerant strategy is beyond the scope of this paper.


{\bf Assumption 1.} The function $g^{(j)}$ is differentiable and convex for all $j=1,\cdots,N$.

{\bf Assumption 2.} The graph $\mathcal{G}$ of the network is undirected and connected.

{\bf Remark 3.} Assumptions 1 and 2 are common in the DO or DOC literature \cite{BG2014,YZZ2017,PL2014,YX2019,YZ2015,YZ2017,ZL2019,SS2018}. In fact, many practical optimization problems can be formalized by the current convex DOC problem or approximated by it using convex relaxation, such as the motion coordination \cite{YX2019}, target aggregation \cite{YZZ2017}, search of radio sources \cite{YZZ2017}, optimal power flow \cite{ED2013}, and so on. The DOC problems for pure-integrator dynamics \cite{PL2014,YX2019,YZ2015}, Euler-Lagrangian systems \cite{YZZ2017} and linear time-invariant systems \cite{YZ2017,ZL2019} have been studied well. This paper further considers more complex nonlinear case (system (1)). Note that a simple combination of traditional DO algorithms \cite{AN2009,IL2011,MZ2012,BJ2008,BG2014} and adaptive backstepping controls \cite{MK1995,HO2017,XD2003,WW2010} will cause the mismatch between (cyber) optimization dynamics and physical dynamics and resultantly cannot guarantee the overall convergence.

Let $y_r^{(j)}$ denote the estimate of agent $\mathcal{D}^{(j)}$ about the value of the solution to (3) and denote $y_r=\mathrm{vec}(y_r^1,\cdots,y_r^N)$ and $L=\mathcal{L}\otimes I_m$. Then problem (3) is equivalent to
\begin{equation}
\min g(y_r)=\sum_{j=1}^Ng^{(j)}(y_r^{(j)}),~\mathrm{subject~to}~Ly_r=0
\end{equation}
Since $g(y_r)$ is convex and the constraint in (4) is linear,
the constrained optimization problem is feasible. The following lemma gives the analysis on the optimal solution of (4).

{\bf Lemma 1 \cite{BG2014}.} Under Assumptions 1 and 2, define by
$$
G(y,v)=g(y)+y^TLv+\frac{1}{2}y^TLy.
$$
Then $G$ is differentiable and convex in its first argument and linear in its second, and:

(i) if $(y^*,v^*)$ is a saddle point of $G$, then $y^*$ is a solution of (4);

(ii) if $y^*$ is a solution of (4), then there exists $v^*$ with $Lv^*=-\nabla g(y^*)$ such that $(y^*,v^*)$ is a saddle point of $G$.

In what follows, we present the resilient DOC scheme by three steps. First, under the healthy conditions, we provide a basic version of DOC protocol (Section 4); Second, under the adversarial conditions, an ADI methodology is proposed to identify the attacked subsystems (Section 5); Third, the final secure DOC scheme is derived by formulating appropriate ADI-based attack countermeasure strategy for the basic version (Section 6).


{\bf Remark 4.} In the cyber superstratum, all the decision-making agents $\mathcal{D}^{(j)}$ themselves are assumed to be healthy in the sense that they will follow any algorithm that we prescribe. However, due to the occurrence of cyber attack $a^{(j)}$, the agent $\mathcal{D}^{(j)}$ and its neighbors will receive false measurements $y^{(j)}$ and thus be induced as ``malicious'' agents in network topology. Moreover, these agents also send false data to other healthy agents and cause cascading failures \cite{OY2015}. These tamped measurements allow the corrupted agents to update their states to arbitrary values such that the corresponding physical subsystems follow false decision commands. In this paper, a resilient DOC algorithm will be developed, where the corresponding secure countermeasure based on an ADI method can effectively avoid the occurrences of ``malicious agents'' and cascading failures.

\section{Basic DOC under healthy environment}

This section deals with the designs of the optimization module $\mathcal{O}^{(j)}$ and control agent $\mathcal{K}^{(j)}$ that form a basic DOC scheme under healthy conditions, i.e., $a^{(j)}(t)=0$ for any $t\ge0$ and $j\in\{1,\cdots,N\}$. In the sequel, we drop the time argument of the signals for notational brevity.

\subsection{DOC control algorithm}

The model of the optimization module $\mathcal{O}^{(j)}$ is designed as the following algorithm
\begin{equation}
\mathcal{O}^{(j)}:\left\{\begin{aligned}
\dot{y}_r^{(j)}=&-\nabla g^{(j)}(y_r^{(j)})-\tilde v^{\mathbf{N}_j}\\
&-(1+\eta)\sum_{i\in \mathbf{N}_j}w_{ji}(y^{(j)}-y^{(i)})\\
\dot v^{(j)}=&\sum_{i\in \mathbf{N}_j}w_{ji}(y^{(j)}-y^{(i)})\\
\Im^{(j)}=&(y_r^{(j)},\nabla g^{(j)}(y_r^{(j)}),\tilde v^{\mathbf{N}_j})
\end{aligned}\right.
\end{equation}
where $\tilde v^{\mathbf{N}_j}=\sum_{i\in \mathbf{N}_j}w_{ji}(v^{(j)}-v^{(i)})\in\mathbb{R}^m$, and $y_r^{(j)}\in \mathbb{R}^m$ and $v^{(j)}\in \mathbb{R}^m$ represent the agent state vectors; $\nabla g^{(j)}$ is the gradient of $g^{(j)}$ and $\eta>0$ is a design parameter, and $\Im^{(j)}\in \mathbb{R}^m\times \mathbb{R}^m\times \mathbb{R}^m$ is the output that serves as a {\bf control command} transmitted to the physical subsystem $\mathcal{P}^{(j)}$.

Next, we present the design of control agent $\mathcal{K}^{(j)}$ which consists of inner-loop module $\mathcal{K}^{(I,j)}$ and outer-loop module $\mathcal{K}^{(O,j)}$ following the control command $\Im^{(j)}$. First, we define the following changes of coordinates
\begin{align}
z_1^{(j)}=x_1^{(j)}-y_r^{(j)},~~z_i=x_i^{(j)}-\alpha_{i-1}^{(j)},~~i=2,\cdots,n
\end{align}
where $\alpha_i^{(j)}=\alpha_{i,I}^{(j)}+\alpha_{i,O}^{(j)}$ is the virtual control function determined at the $i$th step and spitted into two parts: inner-loop control $\alpha_{i,I}^{(j)}$ and outer-loop control $\alpha_{i,O}^{(j)}$. Similarly, $u^{(j)}=u_I^{(j)}+u_O^{(j)}$ is the control input consisting of  inner-loop control $u_I^{(j)}$ and outer-loop control $u_O^{(j)}$. They can be respectively expressed as follows:

\noindent {$\bullet$} Inner-loop control $\mathcal{K}^{(I,j)}(x^{(j)},\Im^{(j)})$
\begin{align}
\nonumber&\alpha_{1,I}^{(j)}=-c_1^{(j)}z_1^{(j)}-\hat\rho^{(j)}z_1^{(j)}-\omega_1^{(j)}\hat\lambda^{(j)}+\hat\pi^{(j)}-S\left(\frac{z_1^{(j)}}{\delta^{(j)}}\right)\\
&\alpha_{i,I}^{(j)}=-z_i^{(j)}-c_i^{(j)}z_i^{(j)}-\omega_i^{(j)}\hat\lambda^{(j)}+\Lambda_i^{(j)}\\
&u_I^{(j)}=\beta_j^{-1}\negthickspace\left[-z_n^{(j)}-c_n^{(j)}z_n^{(j)}-\omega_n^{(j)}\hat\lambda^{(j)}+\Lambda_n^{(j)}\right]
\end{align}
The corresponding update laws are given as
  \begin{align}
\dot{\hat\lambda}^{(j)}&=\Gamma^{(j)}\tau_n^{(j)}\\
\dot\rho^{(j)}&=\gamma_0^{(j)}\|z_1^{(j)}\|^2\\
\dot\pi^{(j)}&=\gamma^{(j)}_1z_1^{(j)}
\end{align}
where
$$
\begin{aligned}
\tau_1^{(j)}&=\omega_1^{(j)}z_1^{(j)},~\tau_i^{(j)}=\tau_{i-1}^{(j)}+\omega_i^{(j)}z_i^{(j)}\\
\omega_i^{(j)}&=\psi_i^{(j)}-\sum_{k=1}^{i-1}\frac{\partial\alpha_{i-2,I}^{(j)}}{\partial x_k^{(j)}}\psi_k^{(j)}\\
\Lambda_i^{(j)}&=\sum_{k=1}^{i-1}\frac{\partial\alpha_{i-1,I}^{(j)}}{\partial x_k^{(j)}}x_{k+1}^{(j)}
+\frac{\partial\alpha_{i-1,I}^{(j)}}{\partial \hat\lambda^{(j)}}\Gamma^{(j)}\tau_i^{(j)}\\
&+\sum_{k=2}^{i-1}\frac{\partial\alpha_{k-1,I}^{(j)}}{\partial\hat\lambda^{(j)}}\Gamma^{(j)} w_i^{(j)}z_k^{(j)}+\frac{\partial\alpha_{i-1,I}^{(j)}}{\partial \hat\pi^{(j)}}\dot{\hat\pi}^{(j)}\\
S\left(\frac{z_1^{(j)}}{\delta^{(j)}}\right)&=\frac{1}{2}\ln\left(1+\frac{z_1^{(j)}}{\delta^{(j)}}\right)
-\frac{1}{2}\ln\left(1-\frac{z_1^{(j)}}{\delta^{(j)}}\right)
\end{aligned}
$$
with $\delta^{(j)}(t)$ being an exponentially decaying function with lower bound $k_b^{(j)}$ such that $|z_{1,s}^{(j)}(0)|<\delta^{(j)}(0)$, and $z_{1,s}^{(j)}$ denotes the $s$th ($s=1,\cdots,m$) element of $z_1^{(j)}$; $\psi_1^{(j)}=\mathrm{diag}\{\varphi_1^{(j)}(x_1^{(j)}),0\}$, $\psi_i^{(j)}=\mathrm{diag}\{\varphi^{(j)}_i(\bar x_i^{(j)}),z_i^{(j)}\}$ for $i=2,\cdots,N$; and $\hat\lambda^{(j)}$, $\hat\rho^{(j)}$ and $\hat\pi^{(j)}$ are the estimates of $\lambda^{(j)}=:[\theta_j^T,\mu]^T$ with $\mu=:((1+\eta)^2\|L\|+\|L\|^3)\Pi^2/2$, $\rho^{(j)}:=(2n-1+\eta)\|L\|$ and $\pi^{(j)}=:\sum_{i\in \mathbf{N}_j}(v^{(j)}_*-v^{(i)}_*)$, respectively, where $\Pi$ is defined in the appendix; $\Gamma^{(j)}$ is a positive definite matrix and $\gamma_0^{(j)}$, $\gamma_1^{(j)}$ and $c_i^{(j)}$ for $i=1,\cdots,n$ are positive constants, all chosen by users. Here the nonlinear function $S(\cdot)$ is introduced to constrain the bound of tracking error $z_1^{(j)}$, motivated by the prescribed performance technique \cite{WW2010}, which will play an important role in enhancing the sensitivity and robustness of the ADI scheme (refer to Remark 7).

\noindent {$\bullet$}  Outer-loop control $\mathcal{K}^{(O,j)}(\Im^{(j)})$
 \begin{align}
 \alpha_{1,O}^{(j)}=&-\nabla g^{(j)}(y_r^{(j)})-2\tilde v^{\mathbf{N}_j}\\
\alpha_{i,O}^{(j)}=&-\frac{\partial\alpha_{i-1,O}^{(j)}}{\partial y_r^{(j)}}\left[\nabla g^{(j)}(y_r^{(j)})+\tilde v^{\mathbf{N}_j}\right]\\
u_O^{(j)}=&-\beta_j^{-1}\frac{\partial\alpha_{n-1,O}^{(j)}}{\partial y_r^{(j)}}\left[\nabla g^{(j)}(y_r^{(j)})+\tilde v^{\mathbf{N}_j}\right]
\end{align}

\begin{figure}
{\small
\vspace{2mm}
\noindent{\bf Algorithm 1:} DOC under healthy environment
\hrule
\hrule
\vspace{0.6mm}
 {\bf DO algorithm (Module $\mathcal{O}(y)$):}
 \begin{equation}
  \begin{aligned}
\dot{y}_r=&-\nabla g(y_r)-Lv-(1+\eta)Ly\\
\dot v=&Ly\\
\Im=&(y_r,\nabla g(y_r),\tilde v)
\end{aligned}
\end{equation}
where $\nabla g(y_r)=\mathrm{vec}(\nabla g^{(1)}(y_r^{(1)}),\cdots,\nabla g^{(N)}(y_r^{(N)}))$ and $\tilde v=\mathrm{vec}(\tilde v^{\mathbf{N}_1},\cdots,\tilde v^{\mathbf{N}_N})$.
\vspace{0.5mm}
\hrule
\vspace{0.5mm}
{\bf Adaptive tracking control (Module $\mathcal{K}(x,\Im)$):}

Inner-loop control $\mathcal{K}^I(x,\Im)$:
  \begin{align}
\alpha_{1,I}=&-C_1z_1-\hat\rho z_1-\omega_1\hat\lambda+\hat\pi-S\left(\frac{z_1}{\delta}\right)\\
\alpha_{i,I}=\nonumber&-z_i-C_iz_i-\omega_i\hat\lambda+\sum_{k=1}^{i-1}\frac{\partial\alpha_{i-1}}{\partial x_k}x_{k+1}\\
&+\frac{\partial\alpha_{i-1}}{\partial \hat\lambda}\Gamma\tau_i+\sum_{k=2}^{n-1}\frac{\partial\alpha_{k-1}}{\partial\hat\lambda}\Gamma w_iz_k+\frac{\partial\alpha_{i-1}}{\partial \hat\pi}\dot{\hat\pi}\\
u_I=\nonumber&B^{-1}\left[-z_n-C_nz_n-\omega_n\hat\lambda+\sum_{k=1}^{n-1}\frac{\partial\alpha_{n-1}}{\partial x_k}x_{k+1}\right.\\
&\left.+\frac{\partial\alpha_{n-1}}{\partial \hat\lambda}\Gamma\tau_n+\sum_{k=2}^{n-1}\frac{\partial\alpha_{k-1}}{\partial\hat\lambda}\Gamma w_iz_k+\frac{\partial\alpha_{n-1}}{\partial \hat\pi}\dot{\hat\pi}\right]
\end{align}
where $C_i=\mathrm{diag}\{c^{(1)}_i,\cdots,c^{(N)}_i\}$, $\hat\rho=\mathrm{diag}\{\hat\rho^{(1)},\cdots,\hat\rho^{(N)}\}$, $\Gamma=\mathrm{diag}\{\Gamma^{(1)},\cdots,\Gamma^{(N)}\}$, $B=\mathrm{diag}\{\beta_1,\cdots,\beta_N\}$, $S(z_1/\delta)=[S(z_1^{(1)}/\delta^{(1)}),\cdots,S(z_1^{(N)}/\delta^{(N)})]$, $\psi_i=\mathrm{diag}\{\psi^{(1)}_i,$ $\cdots,\psi^{(N)}_i\}$, $\omega_i=\mathrm{diag}\{\omega^{(1)}_i,\cdots,\omega^{(N)}_i\}$ and
$$
\begin{aligned}
\tau_1&=\omega_1z_1\\
\tau_i&=\tau_{i-1}+\omega_iz_i\\
\omega_i&=\psi_i-\sum_{k=1}^{i-1}\frac{\partial\alpha_{i-2}}{\partial x_k}\psi_k
\end{aligned}
$$
Outer-loop control $\mathcal{K}^O(\Im)$:
 \begin{align}
 \alpha_{1,O}=&-\nabla g(y_r)-2Lv\\
\alpha_{i,O}=&-\frac{\partial\alpha_{i-1,O}}{\partial y_r}\left[\nabla g(y_r)+Lv\right]\\
u_O=&-B^{-1}\frac{\partial\alpha_{n-1,O}}{\partial y_r}\left[\nabla g(y_r)+Lv\right]
\end{align}
where $i=2,\cdots,n-1$ and $\Gamma_0=\mathrm{diag}\{\gamma^{(1)}_0,\cdots,\gamma^{(N)}_0\}$.
\vspace{0.5mm}
\hrule
\vspace{0.5mm}
{\bf Update laws:}
  \begin{align}
\dot\lambda&=\Gamma\tau_n\\
\dot\rho&=\Gamma_0z_1\circ z_1\\
\dot\pi&=\Gamma_1z_1
\end{align}
where $\Gamma_0=\mathrm{diag}\{\gamma^{(1)}_0,\cdots,\gamma^{(N)}_0\}$ and $\Gamma_1=\mathrm{diag}\{\gamma^{(1)}_1,\cdots,$ $\gamma^{(N)}_1\}$.}
\vspace{2mm}
\hrule
\hrule
\end{figure}

It can be seen that, the DOC structure consists of two-layer dynamics: the optimization dynamics $\mathcal{O}^{(j)}$ and the physical dynamics $(\Sigma^{(j)},\mathcal{K}^{(j)})$ which interact with each other over the communication signals $y^{(j)}$ and $\Im^{(j)}$ (see Fig. 1). Such an architecture also illustrates the CPS's feature that the cyber and physical worlds are integrated. In the inner-loop $\mathcal{K}^{(I,j)}$, a traditional adaptive backstepping controller \cite{MK1995}  (i.e., let $\|L\|=0$) with slight modifications is applied to stabilize the nonlinear strict-feedback system; In the outer-loop $\mathcal{K}^{(O,j)}$, a tracking controller is constructed in order to guarantee that the system output can well track the control command $y_r^{(j)}$ coming from the cyber superstratum. It can be seen that the control laws of the physical systems do not change with the change of the control commands. Summarizing the above procedure (5)-(14), we derive Algorithm 1 for the DOC of the overall CPS under healthy environment.

\subsection{Convergence analysis}

In the section, we discuss the convergence of the proposed DOC algorithm. The main result is stated in the following theorem whose proof is placed in Appendix I.

{\bf Theorem 1.} Under Assumptions 1 and 2, the closed-loop CPS $(\mathcal{P}^{(j)},\mathcal{C}^{(j)})$ with $(\mathcal{K}^{(j)},\mathcal{O}^{(j)})$, $j=1,\cdots,N$ achieves output consensus at an optimal solution $y^\star$ of problem (2), i.e., $\lim_{t\to\infty}y^{(j)}(t)=y^\star$ and all the closed-loop signals are UUB in the absence of cyber attacks if $\eta>2(n-1)$.

{\bf Remark 5.} Differing from the previous works \cite{YZZ2017,PL2014,YX2019,YZ2015,YZ2017,ZL2019} where the integrated closed-loop control laws are designed, this paper presents a new two-layer control structure based on the traditional DO algorithms \cite{AN2009,IL2011,MZ2012,BJ2008,BG2014} and adaptive backstepping controls \cite{MK1995,HO2017,XD2003,WW2010}. Note that the main challenge focuses on how to eliminate the dynamics mismatch between two layers and generate provable optimal consensus. From the proof of Theorem 1 (see (50), (55) and (60)), the dynamics compensation between cyber dynamics $\mathcal{O}^{(j)}$ and physical dynamics $(\Sigma^{(j)},\mathcal{K}^{(j)})$ guarantees the convergence of the overall CPS, where the adaptive mechanism (9)--(11) plays a key role.


\section{Distributed ADI}

This section deals with the design of the monitoring module $\mathcal{M}^{(j)}$, $j\in\{1,\cdots,N\}$. The ADI structure follows the standard one of fault detection and isolation (FDI), formulated by the ARRs of residuals and detection thresholds, e.g., \cite{VR2015,VR2015-2}. However, in this section we will focus on achieving the detection and isolation for {\it arbitrary} malicious behaviors by constructing new residuals and thresholds. Also, due to the coupling effects of multiple propagated attacks on the physical dynamics and optimization dynamics, the design of attack diagnosis becomes more
challenging.

Before giving the main result of this section, we make the following assumption.

{\bf Assumption 3.} The unknown parameter vector $Z:=\mathrm{vec}(y^*,v^*,\theta,\|L\|)$ lies in a known
bounded convex set
$$
\Upsilon_Z=\{Z\in \mathbb{R}^{N(2m+p)+1}:\sigma(Z)\le 0\}
$$
where $\sigma(Z)$ is a convex function.

Assumption 3 is common in the existing results for fault diagnosis \cite{VR2015-2,VR2015,HO2017}, and this is also necessary to detect the attack in transient response phase. It implies that the upper bounds of $y^*,v^*,\theta$ and $\|L\|$, say, $y_M,v_M,\theta_M,L_M$, can be obtained, respectively. Noting that $V(0)$ depends on the unknown vector $Z$, then we define function $\Omega(Z):= V(0)$ and $\bar\Omega:=\sup_{Z\in\Upsilon_Z}\Omega(Z)$, where $V$ is the Lyapunov function defined in proof of Theorem 1.

\subsection{Design of ADI methodology}

The ADI methodology consists of detection filter, adaptive threshold and decision logic. Next, we will give detailed design procedures.

\subsubsection{Detection filter and residual generation}

Now, we design a distributed filter to generate residuals for detecting attacks. According to the dynamics structure (5) of $\mathcal{O}^{(j)}$, the monitoring module $\mathcal{M}^{(j)}$ is designed as
\begin{align}
\mathcal{M}^{(j)}:\left\{\begin{aligned}
\dot{\hat y}_r^{(j)}=&-\nabla g^{(j)}(\hat y_r^{(j)})-\tilde v^{\mathbf{N}_j}\\
&-(1+\eta)\sum_{i\in \mathbf{N}_j}w_{ji}(\hat y_r^{(j)}-y^{(i)})\\
\dot{\hat v}^{(j)}=&\sum_{i\in \mathbf{N}_j}w_{ji}[(\hat y_r^{(j)}-y^{(i)})-(v^{(j)}-\hat v^{(j)})]
\end{aligned}\right.
\end{align}
where $\hat y_r^{(j)}\in \mathbb{R}^m$ and $\hat v^{(j)}\in \mathbb{R}^m$ are the estimates of $y_r^{(j)}$ and $v^{(j)}$ (even $y_r^{(j)}$ and $v^{(j)}$ are available for $\mathcal{M}^{(j)}$), respectively, based on the local communication signals $y^{(i)}$ and $v^{(i)}$, $i\in\{j\}\cup \mathbf{N}_j $. Further, we define two residuals
\begin{align}
e^{(j)}_r&=y^{(j)}_r-\hat y^{(j)}_r\\
e^{(j)}_v&=v^{(j)}-\hat v^{(j)}
\end{align}
Taking (5) and (25) into account, the error dynamics can be expressed as
\begin{align}
\dot e_r^{(j)}=\nonumber&-[\nabla g^{(j)}(y_r^{(j)})-\nabla g^{(j)}(\hat y_r^{(j)})]\\
&-\eta^{(j)}(e_r^{(j)}+z_1^{(j)})-\eta^{(j)}a^{(j)}\\
\dot e_v^{(j)}=&-w_{\mathbf{N}_j}(e_v^{(j)}-e_r^{(j)}-z_1^{(j)})+w_{\mathbf{N}_j}a^{(j)}
\end{align}
where $w_{\mathbf{N}_j}=\sum_{i\in \mathbf{N}_j}w_{ji}$ and $\eta^{(j)}=(1+\eta)w_{\mathbf{N}_j}$.

It is noted that the error dynamics (28)--(29) has a {\it decentralized} form where only own information is used in each error dynamics. The feature means that the coupling effects of the propagated attacks $a^{(i)},i\in\mathbf{N}_j$ on the residuals $e^{(j)}_r$ and $e^{(j)}_v$ caused by the optimization dynamics have been removed such that the locally occurring attack $a^{(j)}$ can be isolated. Later, we will further address the coupling effects of the propagated attacks on the residuals caused by the physical dynamics $z_1^{(j)}$. Moreover, to enhance the attack detectability and remove the existence of stealthy attacks, double coupling residuals have been used here.

If the sensor transmitted information $y^{(j)}$ is not affected by local attack $a^{(j)}$, the error dynamics under healthy conditions, denoted by $(e_{r,H}^{(j)},e_{v,H}^{(j)})$, can be expressed by
\begin{align}
\dot e_{r,H}^{(j)}=\nonumber&-[\nabla g^{(j)}(y_r^{(j)})-\nabla g^{(j)}(\hat y_r^{(j)})]\\
&-\eta^{(j)}(e_{r,H}^{(j)}+z_1^{(j)})\\
\dot e_{v,H}^{(j)}=&-w_{\mathbf{N}_j}(e_{v,H}^{(j)}-e_{r,H}^{(j)}-z_1^{(j)})
\end{align}

The stability of the estimation error dynamics under healthy conditions is analyzed in the following lemma whose proof is placed in Appendix II.

{\bf Lemma 2.} The residuals under the healthy conditions $e_{r,H}^{(j)}(t)$ and $e_{v,H}^{(j)}(t)$ satisfy
\begin{align}
\|e_{r,H}^{(j)}(t)\|\le& e^{-\eta^{(j)}t}e_{r,H}^{(j)}(0)+\Psi(\eta^{(j)},z_1^{(j)}(t),0,t)\\
\|e_{v,H}^{(j)}(t)\|\le\nonumber &e^{-w_{\mathbf{N}_j}t}e_{v,H}^{(j)}(0)\\
&+\Psi(w_{\mathbf{N}_j},e_{r,H}^{(j)}(t)+z_1^{(j)}(t),0,t)
\end{align}
where $\Psi(a,h(t),t_0,t):=a\int_{\tau=t_0}^t e^{a(\tau-t)}\|h(\tau)\|d\tau$.


\subsubsection{Construction of adaptive thresholds}

The $j$th detection thresholds, denoted by $\bar e_{r,H}^{(j)}(t)$ and $\bar e_{v,H}^{(j)}(t)$, are designed based on the bounds of residuals $ e_r^{(j)}(t)$ and $e_v^{(j)}(t)$ under the healthy conditions, respectively. It is noted that the right-hand sides of (32) and (33) cannot be directly used as the thresholds because $z_1^{(j)}=x_1^{(j)}-y_r^{(j)}$($\ne \tilde z_1^{(j)}$) is unavailable for the modules $\mathcal{O}^{(j)}$ and $\mathcal{M}^{(j)}$ due to the existence of cyber attack $a^{(j)}$. To derive an available and reasonable threshold, a heuristic idea is to give a robust design {\it w.r.t.} the unknown ``disturbance-like'' term $z_1^{(j)}$ which, intrinsically, reflects the effects of physical dynamics on the residuals. Hence, we bound the $j$th tracking error $z_1^{(j)}$ under the healthy conditions in the following lemma whose proof is placed in Appendix III.

{\bf Lemma 3.} Under Assumption 3, the servo tracking error $ z_1^{(j)}$ under healthy conditions (i.e., $a^{(j)}=0$) satisfies
$\| z_1^{(j)}(t)\|^2/(2c_1^{(j)})+\int_{\tau=0}^t\| z_1^{(j)}(\tau)\|^2d\tau\le\bar\Omega/c_1^{(j)}$ and $\|z_1^{(j)}(t)\|<\sqrt{m}\delta^{(j)}(t)$.

{\bf Remark 6.} An intuitive method for the ADI design may assess the change of error signal $\tilde z_1^{(j)}:=y^{(j)}-y_r^{(j)}$ based on Lemma 3, because $\tilde z_1^{(j)}=z_1^{(j)}$ under the healthy conditions and $\tilde z_1^{(j)}=z_1^{(j)}+a^{(j)}$ under cyber attacks. However, we emphasize that the error $\tilde z^{(j)}_1$ cannot be directly used as the residual to detect and isolate the cyber attacks because $y_r^{(j)}$ may be simultaneously affected by multiple propagated attacks $a^{(i)},i\in\mathbf{N}_j$ and the locally occurring attack $a^{(j)}$. The adversarial attacker may cooperatively design the stealthy attacks to degrade the system performance while avoiding detection \cite{LW2018,YL2009,YC2019}.

Next, we design the detection threshold based on the bound of $z_1^{(j)}(t)$ under the healthy condition. To be specific, from Lemma 3, one has $z_1^{(j)}(t)\in \Delta^{\delta^{(j)}}_z$ where $\Delta^{\delta^{(j)}}_z:=\{z(t)\in \mathbb{C}^n_m:\frac{1}{2c_1^{(j)}}\|z(t)\|^2+\int_{\tau=0}^t\|z(\tau)\|^2 d\tau\le\frac{\bar\Omega}{c_1^{(j)}},\|z(t)\|\le\sqrt{m}\delta^{(j)}(t)\}$. Substituting the relation into (32) and (33) yields that
$$
\begin{aligned}
\|e_{r,H}^{(j)}(t)\|\le& e^{-\eta^{(j)}t}e_{r,H}^{(j)}(0)\\
&+\sup_{z_1^{(j)}(t)\in\Delta^{\delta^{(j)}}_z}\Psi(\eta^{(j)},z_1^{(j)}(t),0,t)\\
\|e_{v,H}^{(j)}(t)\|\le & e^{-w_{\mathbf{N}_j}t}e_{v,H}^{(j)}(0)\\
&+\sup_{z_1^{(j)}(t)\in\Delta^{\delta^{(j)}}_z}\Psi(w_{\mathbf{N}_j}, e_{r,H}^{(j)}(t)+z_1^{(j)}(t),0,t)
\end{aligned}
$$
Thus, we define the two adaptive thresholds
\begin{align}
\bar e_{r,H}^{(j)}(t)\nonumber&= e^{-\eta^{(j)}t}e_{r,H}^{(j)}(0)+\bar\Psi^{(j)}_{\Delta_z^{\delta^{(j)}}}(\eta^{(j)},0,t)\\
\bar e_{v,H}^{(j)}(t)\nonumber&= e^{-w_{\mathbf{N}_j}t}e_{v,H}^{(j)}(0)+\bar\Psi^{(j)}_{\Delta_{ez}^{\delta^{(j)}}}(w_{\mathbf{N}_j},0,t)
\end{align}
where $\Delta_{ez}^{\delta^{(j)}}:=\{e+z:\|e\|\le\bar e_{r,H}^{(j)},~z\in\Delta_z^{\delta^{(j)}}\}$ and $\bar\Psi^{(j)}_{\Delta^{\delta^{(j)}}}(a,t_0,t):=\sup_{h(t)\in\Delta^{\delta^{(j)}}}a\int_{\tau=t_0}^t e^{a(\tau-t)}\|h(\tau)\|d\tau$.

{\bf Remark 7.} From (28) and (29), multiple propagated
attacks have coupling effects on residuals $e^{(j)}_r$ and $e^{(j)}_v$ over the physical dynamics. To address it, in the inner-loop control module $\mathcal{K}^{(I,j)}$ the modified prescribed performance technique is used to restrict the bound of tracking error $z^{(j)}_1$ (introduce the nonlinear function $S$ into $\alpha_{1,I}^{(j)}$). As a result, the detection thresholds, or further the proposed ADI method, are robust against the multiple propagated attacks. Especially, the prescribed performance bound constraint $\|z^{(j)}_1(t)\|<\sqrt{m}\delta^{(j)}(t)$ is incorporated and contributes to smaller thresholds (from the definition of $\bar\Psi^{(j)}_{\Delta^{\delta^{(j)}}_z}(a,t_0,t)$) and restrain the coupling effects of propagated attacks $a^{(i)},i\in\mathbf{N}_j$ on $e^{(j)}_r$ and $e^{(j)}_v$ such that the sensitivity and isolability to the cyber attacks are improved.

\subsubsection{Decentralized ADI decision logic}

The ADI decision logic implemented in each module $\mathcal{M}^{(j)}$ is based on the ARR, denoted by $\mho^{(j)}(t)$, which is defined as
\begin{align}
\mho^{(j)}(t)=\mho^{(j,r)}(t)\cup \mho^{(j,v)}(t)
\end{align}
where
$$
\begin{aligned}
&\mho^{(j,r)}(t):~\|e_{r,H}^{(j)}(t)\|\le\bar e_{r,H}^{(j)}(t)\\
&\mho^{(j,v)}(t):~\|e_{v,H}^{(j)}(t)\|\le\bar e_{v,H}^{(j)}(t).
\end{aligned}
$$
If $\mho^{(j)}(t)$ is violated, $\mathcal{M}^{(j)}$ will generate an alarm.

The decentralized ADI decision logic is formulated by considering the sensitivity {\it w.r.t} local cyber attacks $a^{(j)}$ and the isolability {\it w.r.t} propagated cyber attacks $a^{(i)}$, $i\in \mathbf{N}_j$, which are summarized in the following theorem.

{\bf Theorem 2.} Consider the ARR $\mho^{(j)}(t)$ defined in (34). The following statements are satisfied:

\begin{description}
  \item[a)]{\bf Attack sensitivity:} If there is a time instant $T_d^{(j)}$ when $\mho^{(j)}(T_d^{(j)})$ is not satisfied, then the occurrence of the local cyber attack $a^{(j)}$ is guaranteed.
  \item[b)]{\bf Attack isolability:} If the transmitted sensor information $y^{(j)}$ is not affected by cyber attack $a^{(j)}$, then the ARR $\mho^{(j)}(t)$ is always satisfied even in the presence of the propagated cyber attacks $a^{(i)}$, $i\in \mathbf{N}_j$.
\end{description}

{\bf Proof.} a) For sake of contradiction, we suppose that no communication attack $a^{(j)}$ has occurred, then $\mho^{(j)}(t)$ is always satisfied according to Lemma 2.

b) Under the condition that $a^{(j)}=0$, even though the propagated cyber attack $a^{(i)}$ may exist, $i\in \mathbf{N}_j$, the estimation error dynamics (28)-(29) reduces to (30)-(31), respectively. Then (32) and (33) are valid and, consequently, $\mho^{(j)}(t)$ is always satisfied.$\hfill{}\blacksquare$

Compared with the existing FDI results \cite{QZ2012,VR2015-2,VR2015}, we have introduced the following techniques to improve the detectability and isolability for attacks:
\begin{itemize}
  \item Double coupling residuals are adopted, which will play a key role in removing stealthy attacks (see Lemma 6).
  \item The modified prescribed performance technique is applied to enhance the sensitivity and isolability to the cyber attacks (See Remark 7).
\end{itemize}


\subsection{Detectability analysis and avoidance of stealthy attacks}

In this section, we will evaluate the attack detectability of the proposed ADI methodology. We first give some properties of functions $\bar\Psi^{(j)}_{\Delta_z^{(j)}}(a,t_0,t)$ and $\bar\Psi^{(j)}_{\Delta_{ez}^{(j)}}(a,t_0,t)$ in the adaptive thresholds, which are important for analyzing the detectability performance of the ADI methodology.

{\bf Lemma 4.} Let $\delta^{(j)}(t)=(k_0^{(j)}e^{-c^{(j)}t}+k_b^{(j)})/\sqrt{m}$, where $k_0^{(j)}$, $k_b^{(j)}$ and $c^{(j)}(\ne a)$ are positive design parameters such that $|z_{1,s}^{(j)}(0)|<\delta^{(j)}(0),~s=1,\cdots,m$. Then

(a) $\bar\Psi^{(j)}_{\Delta^{\delta^{(j)}}_z}(a,0,t)\le
k_b^{(j)}(1-e^{-at})+\frac{ak_0^{(j)}}{a-c^{(j)}}(e^{-c^{(j)}t}-e^{-at})$;

(b) $\bar\Psi^{(j)}_{\Delta^{\delta^{(j)}}_{ez}}(a,0,t)\le
2k_b^{(j)}(1-e^{-at})+\frac{(2a-c^{(j)})ak_0^{(j)}}{(a-c^{(j)})^2}(e^{-c^{(j)}t}-e^{-at})
+a\left[k_b^{(j)}+\frac{ak_0^{(j)}}{a-c^{(j)}}+e_{r,H}^{(j)}(0)\right]te^{-at}$;

(c) $\int_{t=0}^\infty\bar\Psi^{(j)2}_{\Delta^{\delta^{(j)}}_z}(a,0,t)dt\le\bar\Omega/c_1^{(j)}$, $\int_{t=0}^\infty\bar\Psi^{(j)2}_{\Delta^{\delta^{(j)}}_{ez}}(a,0,t)dt\le2\bar\Omega/c_1^{(j)}$.

{\bf Proof.} (a) Note that $\Psi^{(j)}(a,h(t),0,t)$ increases as $\|h(t)\|$ increases. Based on the constraint $\|h(t)\|\le k_0^{(j)}e^{-c^{(j)}t}+k_b^{(j)}$, we have
$$
\bar\Psi^{(j)}_{\Delta^{\delta^{(j)}}_z}(a,0,t)\le a\int_{\tau=0}^t e^{a(\tau-t)}(k_0^{(j)}e^{-c^{(j)}t}+k_b^{(j)})d\tau
$$
By direct computation, the inequality in (a) holds.

(b) Based on (a) and using similar analysis, the proof can be completed.

(c) Let $h^*(t):=\arg\sup_{h(t)\in\Delta^{\delta^{(j)}}_z}a\int_{\tau=0}^t e^{a(\tau-t)}\|h(\tau)\|d\tau$, i.e.,
$\bar\Psi^{(j)}_{\Delta^{\delta^{(j)}}_z}(a,t_0,t)=a\int_{\tau=0}^t e^{a(\tau-t)}\|h^*(\tau)\|d\tau$. Since $\Delta^{\delta^{(j)}}_z$ is a compact set, $h^*(t)$ satisfies $\int_{\tau=0}^\infty\|h^*(\tau)\|^2d\tau\le\bar\Omega/c_1^{(j)}$.

To show (c), we construct the auxiliary dynamics
\begin{equation}
\dot\chi(t)=-a\chi(t)+a\|h^*(t)\|,~\chi(0)=0
\end{equation}
By integrating the dynamics we can find $\chi(t)=\bar\Psi^{(j)}_{\Delta^{\delta^{(j)}}_z}(a,t_0,t)$. On the other hand, considering the Lyapunov function $V=\chi^2/2$, its derivative along with (35) satisfies
$$
\begin{aligned}
\dot V=&\chi(-a\chi+a\|h^*\|)\\
\le&-aV+\frac{a}{2}\|h^*\|^2,
\end{aligned}
$$
integrating two sides of which yields $\int_{t=0}^\infty\chi^2(t)dt\le\bar\Omega/c_1^{(j)}$. Using similar procedure to $\bar\Psi^{(j)}_{\Delta^{\delta^{(j)}}_{ez}}(a,0,t)$, it is easily obtained that $\int_{t=0}^\infty\bar\Psi^{(j)2}_{\Delta^{\delta^{(j)}}_{ez}}(a,0,t)dt\le2\bar\Omega/c_1^{(j)}$.
$\hfill{}\blacksquare$

From Lemma 4-(c), one has $\lim_{t\to\infty}\bar\Psi^{(j)}_{\Delta^{\delta^{(j)}}_z}(a,0,t)=0$ and  $\lim_{t\to\infty}\bar\Psi^{(j)}_{\Delta^{\delta^{(j)}}_{ez}}(a,0,t)=0$ following Barbalat's Lemma.
It means that only if $\mho^{(j)}(t)$ is satisfied, the bound functions $\bar e_{r,H}^{(j)}(t)$ and $\bar e_{v,H}^{(j)}(t)$ will converge to zero, which in turn implies that $e_{r}^{(j)}(t)$ and $e_{v}^{(j)}(t)$ converge to zero. Lemma 4-(a) and -(b) give prescribed performance bounds of $\bar\Psi^{(j)}_{\Delta_z^{\delta^{(j)}}}$ and $\bar\Psi^{(j)}_{\Delta_{ez}^{\delta^{(j)}}}$. By replacing $\bar\Psi^{(j)}_{\Delta_z^{\delta^{(j)}}}$ and $\bar\Psi^{(j)}_{\Delta_{ez}^{\delta^{(j)}}}$ with the  prescribed performance bounds,  we can obtain low-complexity thresholds. However, such relaxations will weaken the detectability and extend the detection time. Also, two modified thresholds converge to $k_b^{(j)}$ and $2k_b^{(j)}$ instead of zero, which may generate the stealthy attacks. Nevertheless, we can choose $k_b^{(j)}$ to be sufficiently small such that the effects of the stealthy attacks resulted from the relaxation are sufficiently small.

To examine the sensitivity of attacks that can be detectable by the proposed attack detection scheme, the following attack detectability is analyzed.

{\bf Lemma 5 (Detectable attacks).} The cyber attack $a^{(j)}$ occurring at the CPS $(\mathcal{P}^{(j)},\mathcal{C}^{(j)})$ is detected using the ARR $\mho^{(j)}$, if there exists some time instant $T_d^{(j)}>T_a^{(j)}$ ($T_a^{(j)}$ is the first time instant of attack $a^{(j)}$ occurrence) such that the attack satisfies
\begin{align}
\nonumber&\eta^{(j)}\left\|\int_{t=T_a^{(j)}}^{T_d^{(j)}} e^{\eta^{(j)}(t-T_d^{(j)})}a^{(j)}(t)dt\right\|>\\
\nonumber&2e^{\eta^{(j)}(T_a^{(j)}-T_d^{(j)})}\|e_r^{(j)}(T_a^{(j)})\|+\bar\Psi^{(j)}_{\Delta_z^{\delta^{(j)}}}(\eta^{(j)},T_a^{(j)},T_d^{(j)})\\
\nonumber&+\eta^{(j)}\int_{t=T_a^{(j)}}^{T_d^{(j)}} e^{\eta^{(j)}(t-T_d^{(j)})}\left\|\nabla g^{(j)}(y_r^{(j)}(t))\right.\\
&\left.-\nabla g^{(j)}(\hat y_r^{(j)}(t))+\eta^{(j)}z_1^{(j)}(t)\right\|dt
\end{align}
or
\begin{align}
\nonumber&w_{\mathbf{N}_j}\left\|\int_{t=T_a^{(j)}}^{T_d^{(j)}} e^{w_{\mathbf{N}_j}(t-T_d^{(j)})}a^{(j)}(t)dt\right\|\\
\nonumber>&2e^{w_{\mathbf{N}_j}(T_a^{(j)}-T_d^{(j)})}\|e_v^{(j)}(T_a^{(j)})\|
+\bar\Psi^{(j)}_{\Delta_{ez}^{\delta^{(j)}}}\negthickspace(w_{\mathbf{N}_j},T_a^{(j)},T_d^{(j)})\\
&+w_{\mathbf{N}_j}\negthickspace\int_{t=T_a^{(j)}}^{T_d^{(j)}} e^{w_{\mathbf{N}_j}(t-T_d^{(j)})}\left\|e_r^{(j)}(t)+z_1^{(j)}(t)\right\|dt
\end{align}
then the attack $a^{(j)}(t)$ is detected at the time $t=T_d^{(j)}$.

{\bf Proof.} After the first occurrence of the attack $a^{(j)}$, i.e., $t>T_a^{(j)}$, the time derivative of $e_r^{(j)}(t)$ becomes
$$
\begin{aligned}
\dot e_r^{(j)}=&-[\nabla g^{(j)}(y_r^{(j)})-\nabla g^{(j)}(\hat y_r^{(j)})]\\
&-\eta^{(j)}(e_r^{(j)}+z_1^{(j)})+\eta^{(j)}a^{(j)}
\end{aligned}
$$
Integrating both sides and applying the triangular inequality yield
$$
\begin{aligned}
\|e_r^{(j)}(T_d^{(j)})\|\ge&\eta^{(j)}\left\|\int_{t=T_f}^{T_d^{(j)}} e^{\eta^{(j)}(t-T_d^{(j)})}a^{(j)}(t)dt\right\|\\
&-e^{\eta^{(j)}(T_a^{(j)}-T_d^{(j)})}\|e_r^{(j)}(T_a^{(j)})\|\\
&-\eta^{(j)}\int_{t=T_a^{(j)}}^{T_d^{(j)}} e^{\eta^{(j)}(t-T_d^{(j)})}\left\|\nabla g^{(j)}(y_r^{(j)}(t))\right.\\
&\left.-\nabla g^{(j)}(\hat y_r^{(j)}(t))+\eta^{(j)}z_1^{(j)}(t)\right\|dt,
\end{aligned}
$$
substituting (36) into which yields
$$
\begin{aligned}
\|e_r^{(j)}(T_d^{(j)})\|>& e^{\eta^{(j)}(T_a^{(j)}-T_d^{(j)})}e_r^{(j)}(T_a^{(j)})\\
&+\bar\Psi^{(j)}_{\Delta_z^{\delta^{(j)}}}(\eta^{(j)},T_a^{(j)},t).
\end{aligned}
$$

Following the similar analysis, (37) guarantees
$$
\begin{aligned}
\|e_v^{(j)}(T_d^{(j)})\|>& e^{w_{\mathbf{N}_j}(T_a^{(j)}-T_d^{(j)})}e_v^{(j)}(T_a^{(j)})\\
&+\bar\Psi^{(j)}_{\Delta_{ez}^{\delta^{(j)}}}(w_{\mathbf{N}_j},T_a^{(j)},t).
\end{aligned}
$$
From the definition of $\mho^{(j)}(t)$, the attack $a^{(j)}(t)$ satisfying (36) or (37) provokes the violation of ARR $\mho^{(j)}(t)$ and resultantly $a^{(j)}(t)$ is detected when $t=T_d^{(j)}$. $\hfill{}\blacksquare$

The inequalities (36)-(37) characterize the class of detectable cyber attacks under the worst-case detectability. The computation of detection time $T_d^{(j)}$ may be somewhat conservative. However, differing from the fault, the attacker may strategically design the (worst-case) attack to extend the detection time as much as possible. Thus, the real-time detection time may sufficiently approach to $T_d^{(j)}$ but not exceed than it.  In general, from (36)-(37), if the cyber attack on the time interval $[T_a^{(j)},T_d^{(j)}]$ is sufficiently large, then the attack can be detected. However, a crafty attacker may ingeniously inject the attack signals which are not detected by the proposed distributed ADI scheme, yet degrade the system performance. The following lemma 6 gives the property of the undetectable attack.

{\bf Lemma 6 (Undetectable attacks).} Suppose that the cyber attack $a^{(j)}(t)$ occurring at the subsystem $(\mathcal{P}^{(j)},$ $\mathcal{C}^{(j)})$ is undetectable by the ARR $\mho^{(j)}$. Then
\begin{align}
\int_{t=T_a^{(j)}}^\infty\left(\int_{\tau=T_a^{(j)}}^t e^{w_{\mathbf{N}_j}(\tau-t)}\|a^{(j)}(\tau)\|d\tau\right)^2 dt\le M
\end{align}
where $M=4\|e_v^{(j)}(T_a^{(j)})\|^2/w_{\mathbf{N}_j}^4+16\bar\Omega/(c_1^{(j)}w_{\mathbf{N}_j}^2)$. Moreover, $\int_{t=T_a^{(j)}}^\infty\|a^{(j)}(t)\|^2dt<+\infty$.

{\bf Proof.} If the attack $a^{(j)}(t)$ occurring at time $T_a^{(j)}$ is not detectable, from Lemma 5, then for any $t\ge T_a^{(j)}$,
\begin{align}
\nonumber&w_{\mathbf{N}_j}\left\|\int_{\tau=T_a^{(j)}}^t e^{w_{\mathbf{N}_j}(\tau-t)}a^{(j)}(\tau)d\tau\right\|\\
\le&2e^{w_{\mathbf{N}_j}(T_a^{(j)}-t)}\|e_v^{(j)}(T_a^{(j)})\|
+2\bar\Psi^{(j)}_{\Delta_{ez}^{\delta^{(j)}}}(w_{\mathbf{N}_j},T_a^{(j)},t)
\end{align}

Consider the right-hand side of (39). Taking square and integral consecutively to each term yields
$$
\begin{aligned}
&4\|e_v^{(j)}(T_a^{(j)})\|^2\int_{t=T_a^{(j)}}^\infty e^{2w_{\mathbf{N}_j}(T_a^{(j)}-t)} dt\le\frac{2\|e_v^{(j)}(T_a^{(j)})\|^2}{w_{\mathbf{N}_j}},\\
&4\int_{t=T_a^{(j)}}^\infty\bar\Psi^{(j)2}_{\Delta_{ez}^{\delta^{(j)}}}(w_{\mathbf{N}_j},T_a^{(j)},t)dt
\le\frac{8\bar\Omega}{c_1^{(j)}}.
\end{aligned}
$$
where the second inequality follows from Lemma 4-(c).

Then using the Cauchy-Buniakowsky-Schwarz inequality, one has
\begin{align}
\nonumber&4\int_{t=T_a^{(j)}}^\infty(e^{w_{\mathbf{N}_j}(T_a^{(j)}-t)}\|e_v^{(j)}(T_a^{(j)})\|\\
\nonumber&+\bar\Psi^{(j)}_{\Delta_{ez}^{\delta^{(j)}}}(w_{\mathbf{N}_j},T_a^{(j)},t))^2dt\\
&\le\frac{4\|e_v^{(j)}(T_a^{(j)})\|^2}{w_{\mathbf{N}_j}^2}+\frac{16}{c_1^{(j)}}\bar\Omega
\end{align}
Combining (39) and (40), Eq. (38) follows at once. Further, $\lim_{t\to\infty}\int_{\tau=T_a^{(j)}}^t e^{w_{\mathbf{N}_j}(\tau-t)}a^{(j)}(\tau)d\tau=0$.

Next, to prove $\int_{t=T_a^{(j)}}^\infty\|a^{(j)}(t)\|^2dt<+\infty$, we consider the error dynamics
$$
\dot e_v^{(j)}=-w_{\mathbf{N}_j}(e_v^{(j)}-e_r^{(j)}-z_1^{(j)})+w_{\mathbf{N}_j}a^{(j)}.
$$
Noting that $\mho^{(j)}$ is always satisfied, then $e_r^{(j)},e_v^{(j)}, z_1^{(j)}\in L_2$ from Lemma 4-(c). Therefore, there exist a sufficiently big $T\ge T_a^{(j)}$ and a time interval $\Xi_v$ with $\nu(\Xi_v)=0$  such that
$$
\frac{a^{(j)}_s(t)}{e_{v,s}^{(j)}(t)}<1,~\forall t\in [T,\infty)\backslash\Xi_v
$$
which means that there exists a function $\bar\phi_v(t)\le 0$ such that
\begin{align}
\|a^{(j)}(t)\|\negthickspace<\|e_v^{(j)}(t)\|~\mathrm{or}~a^{(j)}(t)=\bar\phi_v(t)\mathrm{sgn}(e_v^{(j)}(t))
\end{align}
for any $t\in [T,\infty)\backslash\Xi_v$, where $a_s^{(j)}$ and $e^{(j)}_{v,s}$ represent the $s$th element of $a^{(j)}$ and $e^{(j)}_v$. Applying similar procedure to $
\dot e_r^{(j)}=-[\nabla g^{(j)}(y_r^{(j)})-\nabla g^{(j)}(\hat y_r^{(j)})]-\eta^{(j)}(e_r^{(j)}+z_1^{(j)})+\eta^{(j)}a^{(j)}
$, there exist $\bar\phi_r(t)\le 0$ and $\Xi_r$ with $\nu(\Xi_r)=0$ such that
\begin{align}
\|a^{(j)}(t)\|\negthickspace<\|e_r^{(j)}(t)\|~\mathrm{or}~a^{(j)}(t)=\bar\phi_r(t)\mathrm{sgn}(e_r^{(j)}(t))
\end{align}
for any $t\in [T,+\infty)\backslash\Xi_r$.

Compared (41) with (42), and noting that the equality
$$
\mathrm{sgn}(e_v^{(j)}(t))=\mathrm{sgn}(e_r^{(j)}(t)),\forall t\in[T,+\infty)\backslash(\Xi_r\cup\Xi_v)
$$
does not hold, it yields that $\|a^{(j)}(t)\|<\|e_r^{(j)}(t)\|$ or $\|a^{(j)}(t)\|<\|e_v^{(j)}(t)\|$ for any $t\in [T,+\infty)\backslash(\Xi_r\cup\Xi_v)$, which guarantees $\int_{t=T_a^{(j)}}^\infty\|a^{(j)}(t)\|^2dt<+\infty$. $\hfill{}\blacksquare$



Lemma 6 implies that any undetectable attack must belong to $L_2$. From its proof, we can see the design of double coupling residuals plays a key role in removing the existence of stealthy attacks $a^{(j)}(t)=\bar\phi_r(t)\mathrm{sgn}(e_r^{(j)}(t))$ against $\mho^{(j,r)}$ or $a^{(j)}(t)=\bar\phi_v(t)\mathrm{sgn}(e_v^{(j)}(t))$ against $\mho^{(j,v)}$. Now, we give the main result of this section.

{\bf Theorem 3.} Under Assumptions 1-3, the closed-loop CPS $(\mathcal{P}^{(j)},\mathcal{C}^{(j)})$ with $(\mathcal{K}^{(j)},\mathcal{D}^{(j)}(\mathcal{O}^{(j)},$ $\mathcal{M}^{(j)}))$ achieves output consensus at an optimal solution of problem (2) and all the closed-loop signals are UUB even in the presence of the undetectable attacks if $\eta>2(n-1)$.


{\bf Proof.} From Lemma 6, one has $\int_{t=T_a^{(j)}}^\infty\|a^{(j)}(t)\|^2dt<+\infty$. Following Theorem 1, the proof can be complete. $\hfill{}\blacksquare$

Theorem 3 implies that all the subsystems $(\mathcal{P}^{(j)},\mathcal{C}^{(j)})$, $j=1,\cdots,N$ can achieve optimal consensus only if the ARRs $\mho^{(j)}$ are satisfied. In other words, {\it any} attacks cannot bypass the designed ADI methodology to destroy the system convergence.


\section{Secure countermeasure against cyber attacks}

With these results on basic DOC and ADI in hand, we now provide a secure countermeasure against the cyber attacks and give the final secure DOC algorithm. The {\bf security objective} is to steer the physical part $\mathcal{P}^{(j)}$, $j\in \mathbf{N}_A$ to a secure state $y_s^{(j)}\in \mathbb{R}^m$, i.e., $\lim_{t\to\infty}y^{(j)}(t)=y_s^{(j)}$, while guaranteeing $(\mathcal{P}^{(j)},\mathcal{C}^{(j)})$, $j\in \mathbf{N}_H$ to achieve the output consensus at the optimal solution of
\begin{equation}
\min\sum_{j\in \mathbf{N}_H}g^{(j)}(s),~s\in \mathbb{R}^m
\end{equation}
where $\mathbf{N}_A$ represents the set of subsystems $(\mathcal{P}^{(j)},\mathcal{C}^{(j)})$ which are affected by detectable attack $a^{(j)}$ subject to (36) or (37), and $\mathbf{N}_H\triangleq\{1,\cdots,N\}\setminus\mathbf{N}_A$ represents the set of the subsystems which are healthy or affected by undetectable attacks satisfying (38). To guarantee the output consensus of subsystems $\mathcal{P}^{(j)}$, $j\in \mathbf{N}_H$, the following assumption is necessary in accordance with Assumption 2.

{\bf Assumption 4.} The network topology induced by agents $\mathcal{D}^{(j)}$, $j\in \mathbf{N}_H$ is connected.

Assumption 4 captures the communication redundancy of graph $\mathcal{G}$. Note that different notions of network robustness have been reported to guarantee the convergence of resilient distributed algorithms, e.g., \cite{LS2015,SS2018,CZ2019,WF2019}. For an undirected graph, Assumption 4 is in fact necessary for achieving the security objective (43).

Before giving the secure countermeasure, we first define a notification signal $\digamma^{(j)}(t)$ such that ``$\digamma^{(j)}(t)=1$'' represents the $j$th subsystem $(\mathcal{P}^{(j)},\mathcal{C}^{(j)})$ is attacked at time $t$, and ``$\digamma^{(j)}(t)=0$'' otherwise. In order to prevent the transmission data $y^{(j)}$ corrupted by the cyber attack $a^{(j)}$ from being propagated the neighboring subsystems, we design
\begin{equation}
\digamma^{(j)}(t)=\left\{
\begin{aligned}
&1,~\mathrm{if}~t\ge T_d^{(j)}\\
&0,~\mathrm{otherwise}
\end{aligned}\right.
\end{equation}
where $T_d^{(j)}$ is the attack detection time for $\mathcal{M}^{(j)}$, defined as
$$
T_d^{(j)}=\inf_{t\ge 0}\left\{t:\mho^{(j)}(t)~\mathrm{is~volated}\right\}.
$$
If $\mho^{(j)}(t)$ is always satisfied, then the detection time is defined as $T_d^{(j)}=+\infty$.

According to the security objective, we modify the output of decision-making dynamics (5), i.e., the control command $\Im^{(j)}$, under adversarial environment as
\begin{equation}
\Im^{(j)}=\left\{
\begin{aligned}
&(y_r^{(j)},\nabla g^{(j)}(y_r^{(j)}),\tilde v^{\mathbf{N}_j}),~\mathrm{if}~t<T_d^{(j)}\\
&(y_s^{(j)},0,0),~~~~~~~~~~~~~~~~~\mathrm{otherwise}
\end{aligned}\right.
\end{equation}

The final {\bf secure decision-making algorithm} for agent $\mathcal{D}^{(j)}$ based on the notification signal (44) and the control command (45) is summarized as:

$\bullet$ Receive $(y^{(i)},v^{(i)},\digamma^{(i)})$ to its neighbors  $\mathcal{D}^{(i)},i\in \mathbf{N}_j$; \\
$\bullet$ Set $y^{(i)}=0$ and $v^{(i)}=0$ if $\digamma^{(i)}=1$;\\
$\bullet$ Update state by computing Eq. (5);\\
$\bullet$ Send control command $\Im^{(j)}$ to control module $\mathcal{K}^{(j)}$

{\bf Theorem 4.} Consider the closed-loop CPS $(\mathcal{P}^{(j)},\mathcal{C}^{(j)})$ with $(\mathcal{K}^{(j)},\mathcal{D}^{(j)}(\mathcal{O}^{(j)},\mathcal{M}^{(j)}))$ in the presence of cyber attacks $a^{(j)}$, $j\in\{1,\cdots,N\}$. Under Assumptions 1-4, subsystems $(\mathcal{P}^{(j)},\mathcal{C}^{(j)})$, $j\in \mathbf{N}_H$ achieve the output consensus at the optimal solution $y^\star_{\mathrm{N}_H}$ of (43), while the system output of physical part $\mathcal{P}^{(j)}$, $j\in \mathbf{N}_A$ converges to a given state $y_s^{(j)}$, i.e., $\lim_{t\to\infty} y^{(j)}(t)=y^\star_{\mathrm{N}_H}$ for $j\in \mathbf{N}_H$ and $\lim_{t\to\infty}y^{(j)}(t)=y_s^{(j)}$ for $j\in \mathbf{N}_A$. Moreover, all the closed-loop signals are UUB.

{\bf Proof.} Consider the subsystem $(\mathcal{P}^{(j)},\mathcal{C}^{(j)})$, $j\in \mathbf{N}_A$. From Lemma 5, $\mho^{(j)}$ is not satisfied and the optimization module $\mathcal{O}^{(j)}$ sends the control command $(y_s^{(j)},0,0)$ to the control module $\mathcal{K}^{(O,j)}$. Then from (12)-(14) the outer-loop control $u^{(j)}_O=0$ and the inner-loop control $u^{(j)}_I$ (traditional adaptive backstepping control \cite{MK1995}) can guarantee the closed-loop $\mathcal{P}^{(j)}$ converges to $y_s^{(j)}$.

Consider the subsystem $(\mathcal{P}^{(j)},\mathcal{C}^{(j)})$, $j\in \mathbf{N}_H$. Given the above secure decision-making, the dynamics (5) becomes
$$
\left\{\begin{aligned}
\dot{y}_r^{(j)}=&-\nabla g^{(j)}(y_r^{(j)})-\sum_{i\in \mathbf{N}_j\cap\mathbf{N}_H}w_{ji}(v^{(j)}-v^{(i)})\\
&-(1+\eta)\sum_{i\in \mathbf{N}_j\cap\mathbf{N}_H}w_{ji}(y^{(j)}-y^{(i)})\\
\dot v^{(j)}=&\sum_{i\in \mathbf{N}_j\cap\mathbf{N}_H}w_{ji}(y^{(j)}-y^{(i)})
\end{aligned}\right.
$$

Following Theorem 1 and Theorem 3, the output of $\mathcal{P}^{(j)}$, $j\in\mathbf{N}_H$ converges to the optimal solution of problem (43) as $t\to+\infty$, and all the signals are UUB. $\hfill{}\blacksquare$

From (44) and (45), the security performance under the cyber attacks heavily relies on the detection time $T_d^{(j)}$. With the increase of $(T_d^{(j)}-T_a^{(j)})$, the attacker will have more time to damage the system performance.

\begin{figure}
  \centering
  \includegraphics[height=4.5cm,width=8cm]{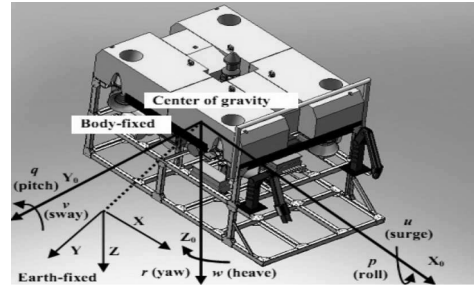}\\
  \caption{Model of Underwater Robotics Vehicle}
\end{figure}

\section{Simulation example}

As a practical application of the studied problem framework, we apply our algorithms to the problem of motion coordination of multiple Remotely Operated Vehicles (ROVs). The motion coordination expects the formation of ROVs to rendezvous at a location which is optimal
for the formation \cite{YZ2017,YX2019}. The dynamic behavior of ROVs can be described in two coordinate frames, the body-fixed frame and the earth-fix
frame as shown in Fig. 2. The dynamics equation of each ROV
can be expressed as \cite{KZ2011}:
\begin{equation}
\begin{aligned}
\dot{\boldsymbol\eta}=&J(\boldsymbol\eta)\boldsymbol\nu\\
M\dot{\boldsymbol\nu}+C(\boldsymbol\nu)+D(\boldsymbol\nu)\boldsymbol\nu+g(\boldsymbol\eta)=&\boldsymbol\tau+\Delta\boldsymbol f
\end{aligned}
\end{equation}
where $\boldsymbol\eta=[x,y,z,\phi,\theta,\psi]^T$ is the position and orientation described in the earth-fixed frame ($|\theta|<\pi/2$ and $|\phi|<\pi/2$),  $\boldsymbol\nu=[u,v,w,p,q,r]^T$ is the linear and angular velocity in the body-fixed frame, $M=M_{RB}+M_A$ and $M$ is positive definite, $C(\boldsymbol\nu)=C_{RB}(\boldsymbol\nu)+C_A(\boldsymbol\nu)$ satisfying $C(\boldsymbol\nu)=-C^T(\boldsymbol\nu)$, $M_{RB}$ is the rigid-body inertia matrix, $M_A$ is the added inertia matrix;
$C_{RB}(\boldsymbol\nu)$ is the rigid-body Coriolis and centripetal matrix,
$C_A(\boldsymbol\nu)$ is the hydrodynamic Coriolis and centripetal matrix
in cluding added mass, $D(\boldsymbol\nu)$ is hydrodynamic damping
and lift matrix, $g(\boldsymbol\eta)$ is a vector of gravitational forces and
moment, $\boldsymbol\tau$ is the control force and torque vector, $\Delta\boldsymbol f$ is the bounded disturbance vector. Note that system (46) can be transformed into the form of system (1) by choosing the state variables $[x_1^T,x_2^T]^T=[\boldsymbol\eta^T,\boldsymbol\nu^TJ^T(\boldsymbol\eta)]^T$.

As reported in \cite{KZ2011}, in the positioning and trajectory tracking control of ROV, the variables needed to be controlled are $x,y,z$ and $\psi$. Under some cases, for the purpose of improving the dynamic stability and decreasing the influences of $\phi$ and $\theta$ on other variables, a simple P-controller can be used to control $\phi$ and $\theta$. Therefore the order of the MIMO backstepping robust controller can be reduced from 6 degrees of freedom (DOF) to 4 DOF.

To simplify the controller design, the transformation
matrix $J(\boldsymbol\eta)$ can also be approximately obtained by assuming that $\phi=\theta=p=q=0$, then the corresponding matrix parameters of reduced system are $M=\mathrm{diag}\{m_\nu-X_{\dot u},m_\nu-Y_{\dot v},m_\nu-Z_{\dot w},I_z-N_{\dot r}\}$, $
D(\boldsymbol\nu)=-\mathrm{diag}\{X_u+X_{u|u|},Y_v+Y_{v|v|}v,Z_w+Z_{w|w|},N_r+N_{r|r|}r\}$, $g(\eta)=[0,0,-(W-B),0]^T$ and
$$
\begin{aligned}
J(\boldsymbol\eta)=&\left[\begin{matrix}
\cos\psi & -\sin\psi  & 0 & 0\\
\sin\psi & \cos\psi  & 0 & 0\\
0 & 0   &  1 & 0\\
0 & 0   &  0 & 1\\
\end{matrix}\right],\\
C(\boldsymbol\nu)=&\left[\begin{matrix}
0 & 0  & 0 & -(m_\nu-Y_{\dot v})v\\
0 & 0  & 0 & -(m_\nu-X_{\dot u})u\\
0 & 0   &  0 & 0\\
(m_\nu-Y_{\dot v})v & -(m_\nu-X_{\dot u})u  &  0 & 0\\
\end{matrix}\right].
\end{aligned}
$$
According to \cite{KZ2011}, the velocity dynamics can be expressed as linear-parametric form
$$
M\dot{\boldsymbol\nu}_v+C(\boldsymbol\nu)+D(\boldsymbol\nu)\boldsymbol\nu+g(\boldsymbol\eta)=\Phi^T(\boldsymbol\nu,\dot{\boldsymbol\nu}_v,\boldsymbol\eta)\sigma
$$
where  $\sigma=[m_\nu-X_{\dot u},m_\nu-Y_{\dot v},X_u,X_{|u|u},Y_v,Y_{|v|v},m_{\nu}-Z_{\dot w},Z_w,Z_{w|w|},W-B,I_z-N_{\dot r},N_r,N_{r|r|}]$ is unknown system parameter vector, $\boldsymbol\nu_v$ is the virtual control and $\Phi(\boldsymbol\nu,\dot{\boldsymbol\nu}_v,\boldsymbol\eta)$ is a known reduced regressor matrix function whose specific form can be found in \cite{KZ2011} and is omitted here for saving space.

 \begin{table}[!h]
\renewcommand{\arraystretch}{1.2}
\center{{\bf TABLE I.} Simulation Model Parameters of ROV}
\begin{center}
\def\temptablewidth{0.48\textwidth}
{\rule{\temptablewidth}{1pt}}
\begin{tabular*}{\temptablewidth}{@{\extracolsep{\fill}}ll|ll}
Par                       & Value      & Par   &   Value                     \\   \hline
$m_\nu$  & 2500kg    & $X_{\dot u}$   &   -2140kg             \\
$I_z$  & 1250kg$\cdot$m$^2$    & $Y_{\dot v}$  &   -1636kg           \\
$W$    & 24525N    & $Z_{\dot w}$           &-3000kg      \\
$B$    & 24525N           &$N_{\dot r}$       &-1524kg$\cdot$m$^2$    \\
$X_u$  & -3610kg/s                & $X_{u|u|}$   &   -952 kg/m      \\
$Y_v$  & -4660kg/s                 & $Y_{v|v|}$  &-1361kg/m                  \\
$Z_w$  & -11772kg/s                 & $Z_{w|w|}$ & -3561kg/m\\
$N_r$  & -7848kg$\cdot$m$^2$/(s$\cdot$rad) & $N_{r|r|}$ &-773kg$\cdot$m$^2$/(s$\cdot$rad)                       \\
\end{tabular*}
{\rule{\temptablewidth}{1pt}}
\end{center}
\end{table}

Consider a ROV formation which consists of 4 same ROVs.
The parameters of the ROV are shown in Table I. The communication topology $\mathcal{G}$ is given by a $2$-regular graph and the edge weight $w_{ji}=1$. The problem of multi-agent coordination consisting in finding a distributed control strategy that is able to drive each ROV from its initial position to rendezvous at the target position which minimizes the square sum of distances from these initial positions. The coordination control objective can be formulated as the following problem:
\begin{equation}
\min_{\boldsymbol\eta^{(j)}} \sum_{j=1}^4\|\boldsymbol\eta^{(j)}-\boldsymbol\eta^{(j)}_0\|^2,~\mathrm{s.t.}~\boldsymbol\eta^{(1)}=\cdots=\boldsymbol\eta^{(4)}
\end{equation}
where $\boldsymbol\eta^{(j)}_0$ represents the initial state of the $j$th ROV.

Next, we apply the proposed secure DOC control strategy to complete the motion coordination task. Consider the cyber attacks (also including the sensor faults or some extraneous factors such as ocean currents) occurring in the complex underwater environment. When the cyber core detects the existence of the cyber attacks, it will drive the attacked ROV to the secure state $\boldsymbol\eta_s=0$.
In the simulation, the initial state conditions of these four ROVs are set as $\boldsymbol\eta^{(1)}(0)=[0.3~0.4~1~0]^T$, $\boldsymbol\eta^{(2)}(0)=[0.1~ 0.1~ 0.5~ -\pi/6]^T$, $\boldsymbol\eta^{(3)}(0)=[0~ 0~ 0~ -\pi/8]^T$ and $\boldsymbol\eta^{(4)}(0)=[0.2~ 0.5~ 1~ 0]^T$. Assume that the 4th ROV suffers the cyber attack at $t=30$s, and $\phi^{(4)}(t)=e^{0.5(t-30)-1}[\sin(t)~\cos(t)~-\sin(t)~-\cos(t)]^T$. For simplifying calculation, only the ARR $\mho^{(j,r)}(t)$ rather than $\mho^{(j)}(t)$ is used in the proposed ADI approach.

\begin{figure}
  \centering
  \includegraphics[height=3.9cm,width=8cm]{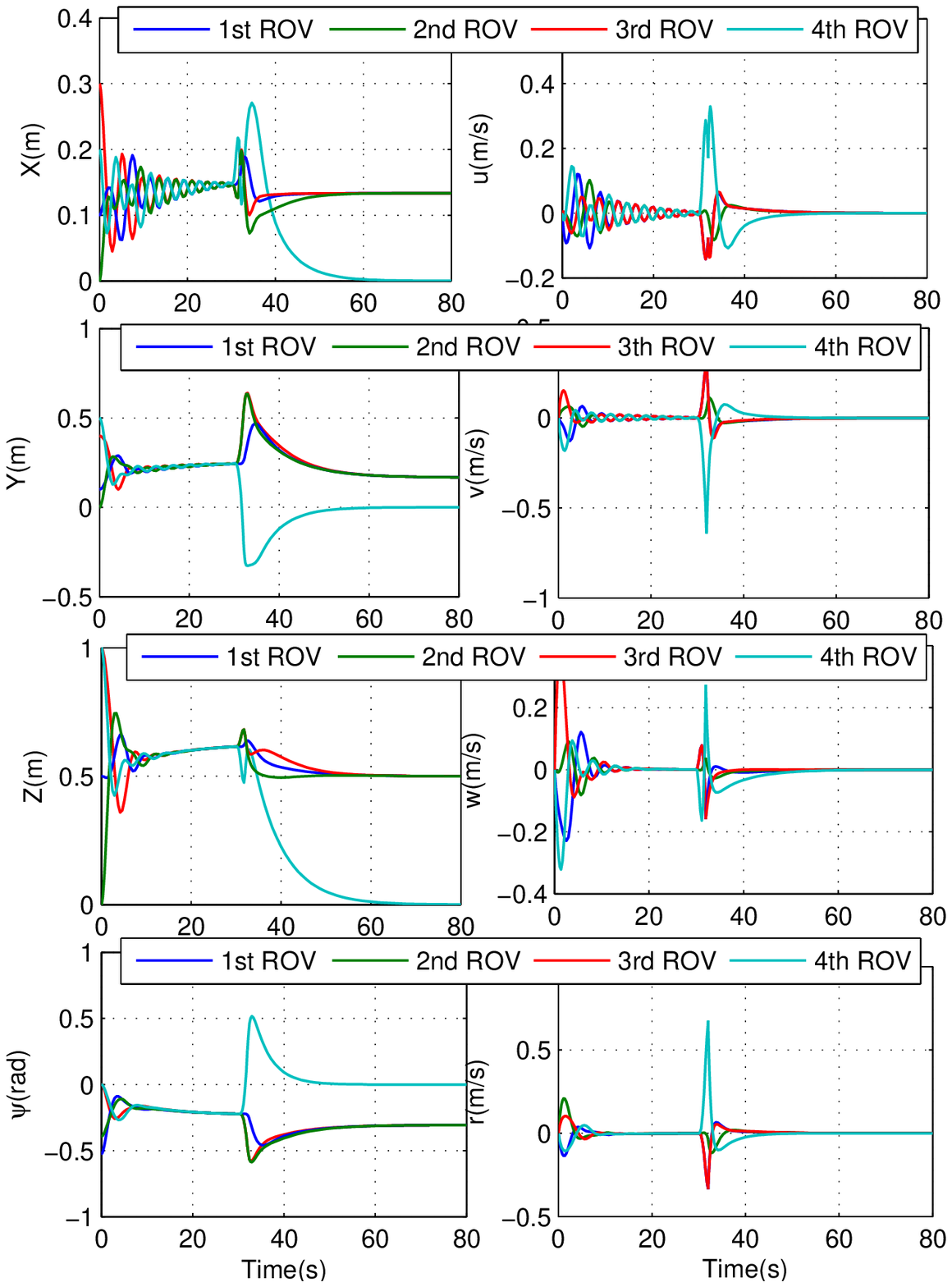}\\
  \caption{Trajectories of $\boldsymbol\eta(t)$ and $\boldsymbol\nu(t)$ of four ROVs.}
  \centering
  \includegraphics[height=3.9cm,width=8cm]{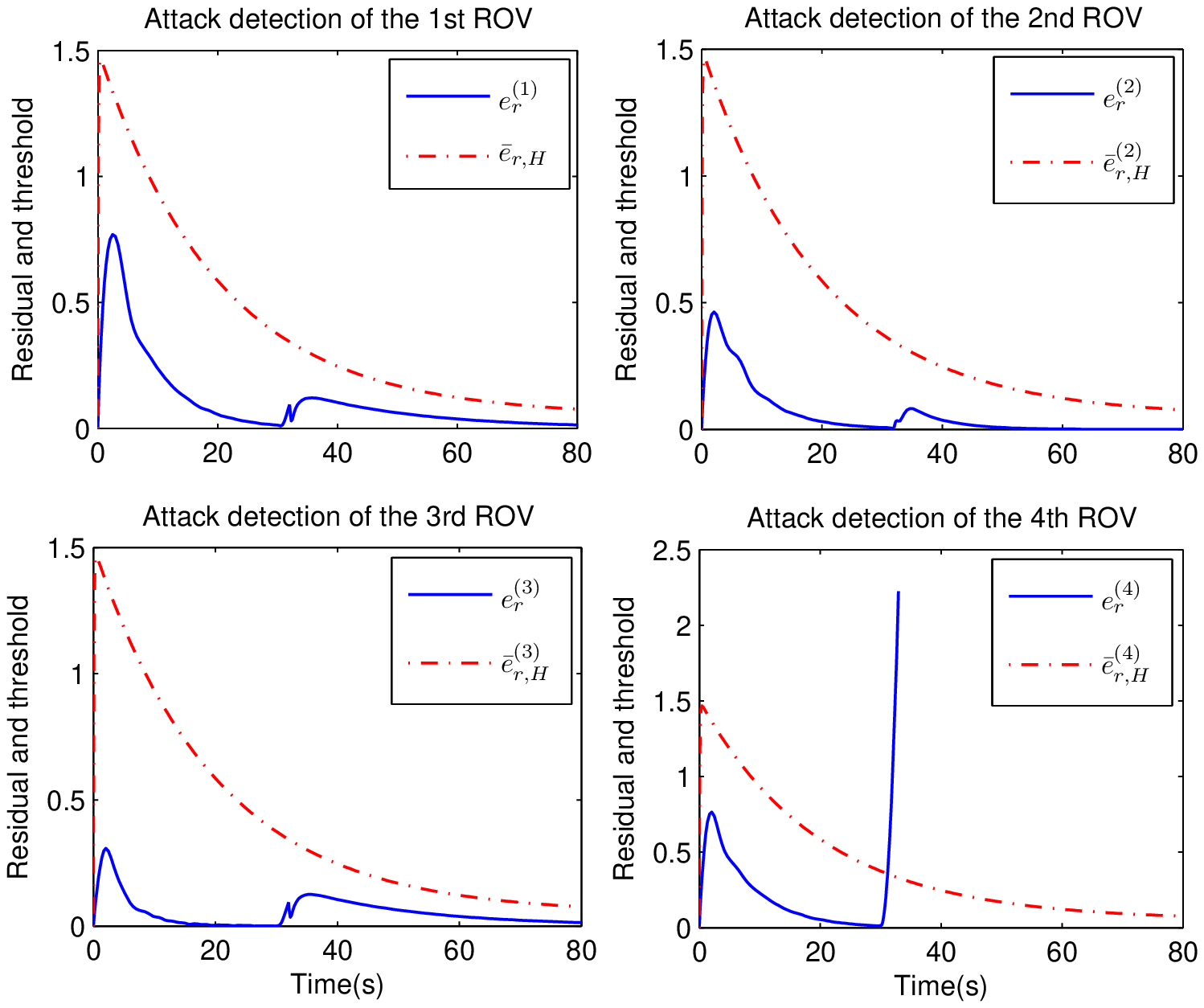}\\
  \caption{ADI by the ARR $\mho^{(j,r)}(t)$, $j=1,\cdots,4$.}
  \centering
  \includegraphics[height=3.9cm,width=8cm]{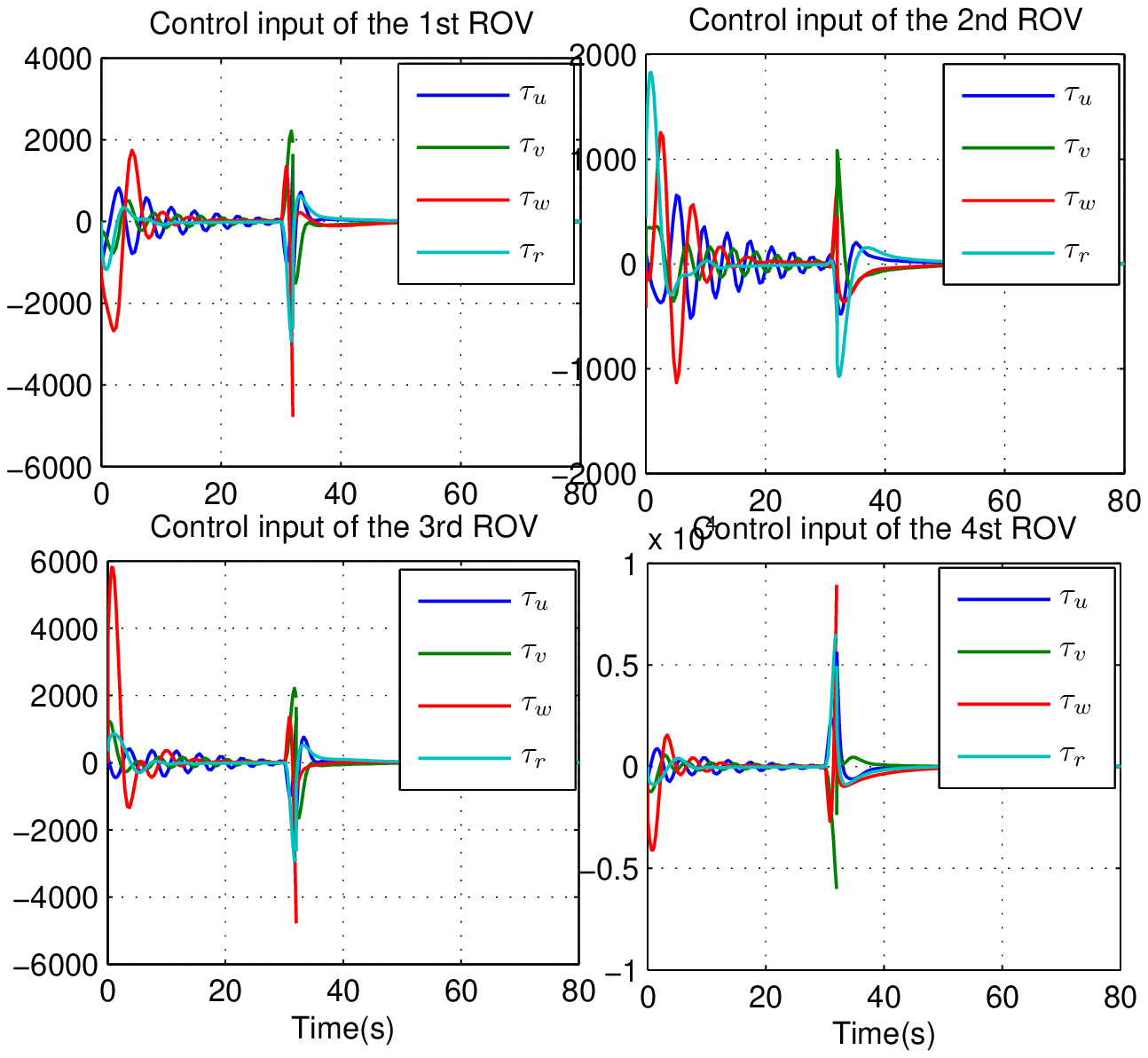}\\
  \caption{Control inputs of four ROVs ($\tau_u$:kgf, $\tau_v$:kgf, $\tau_w$:kgf, $\tau_r$:kgf$\cdot$m).}
\end{figure}

\begin{figure}
  \centering
  \includegraphics[height=3.9cm,width=8cm]{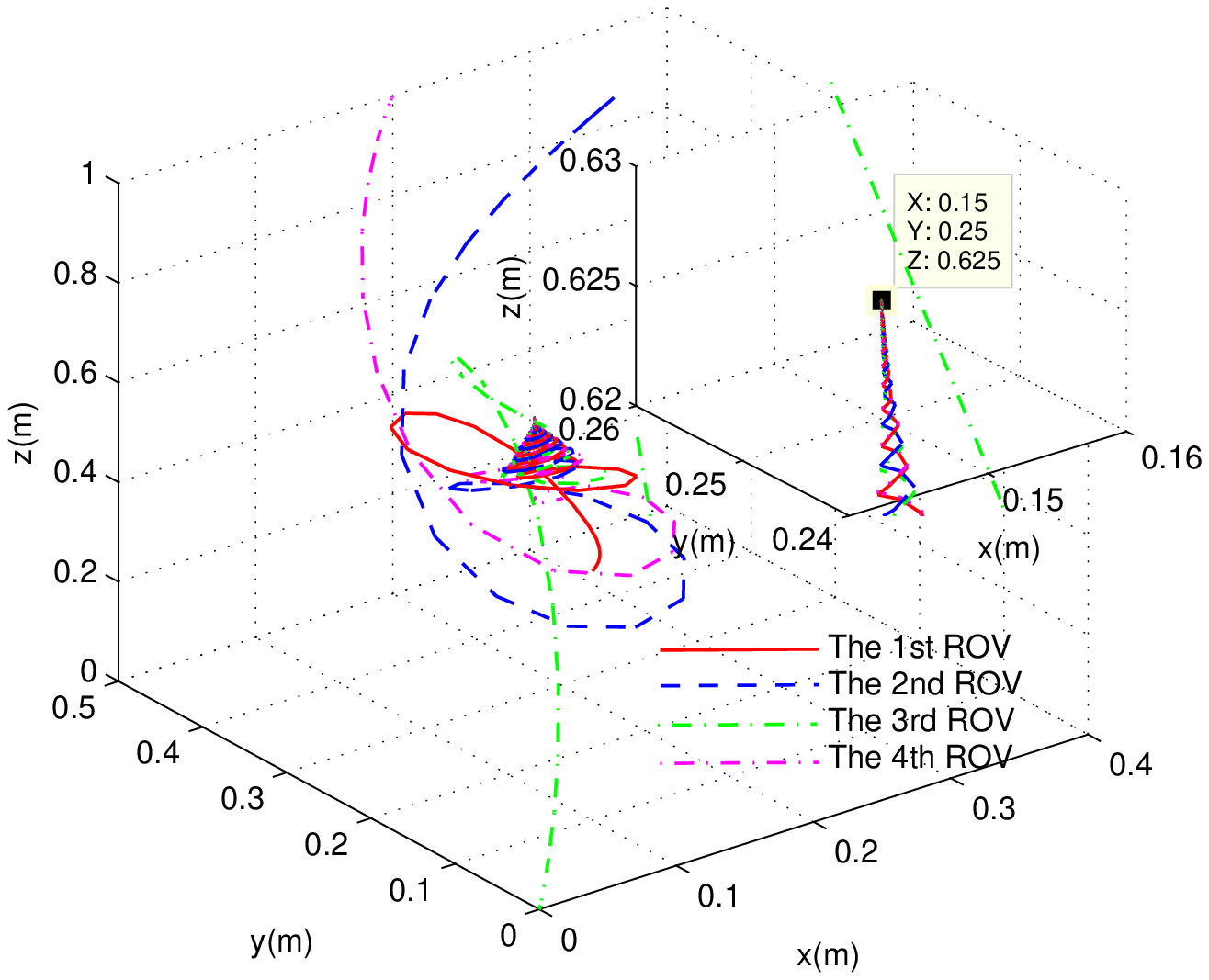}\\
  \caption{3-dimension routes of four ROVs on [0s,80s] under Case 1.}
  \centering
  \includegraphics[height=3.9cm,width=8cm]{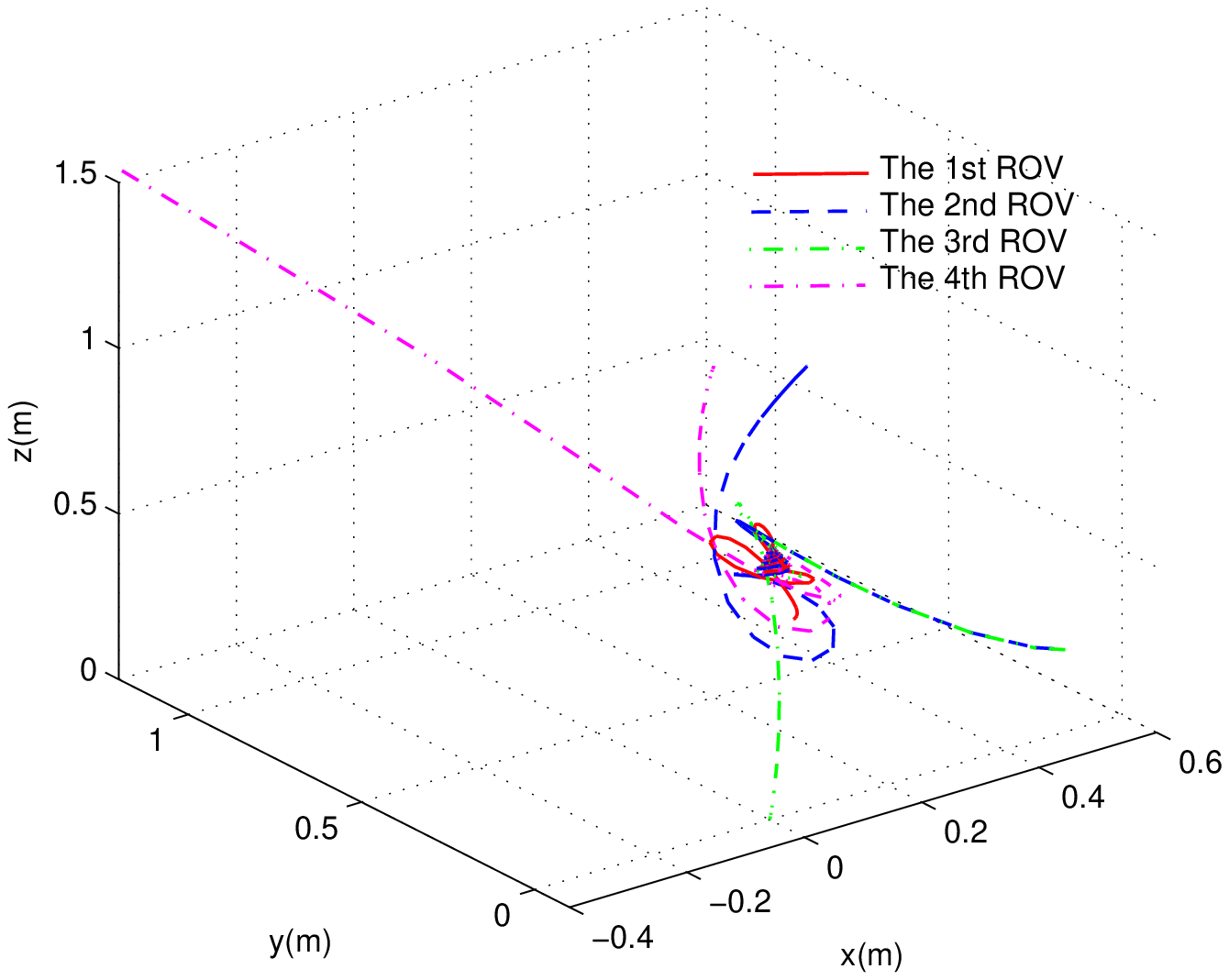}\\
  \caption{3-dimension routes of four ROVs on [0s,34.45s] under Case 2.}
  \centering
  \includegraphics[height=3.9cm,width=8cm]{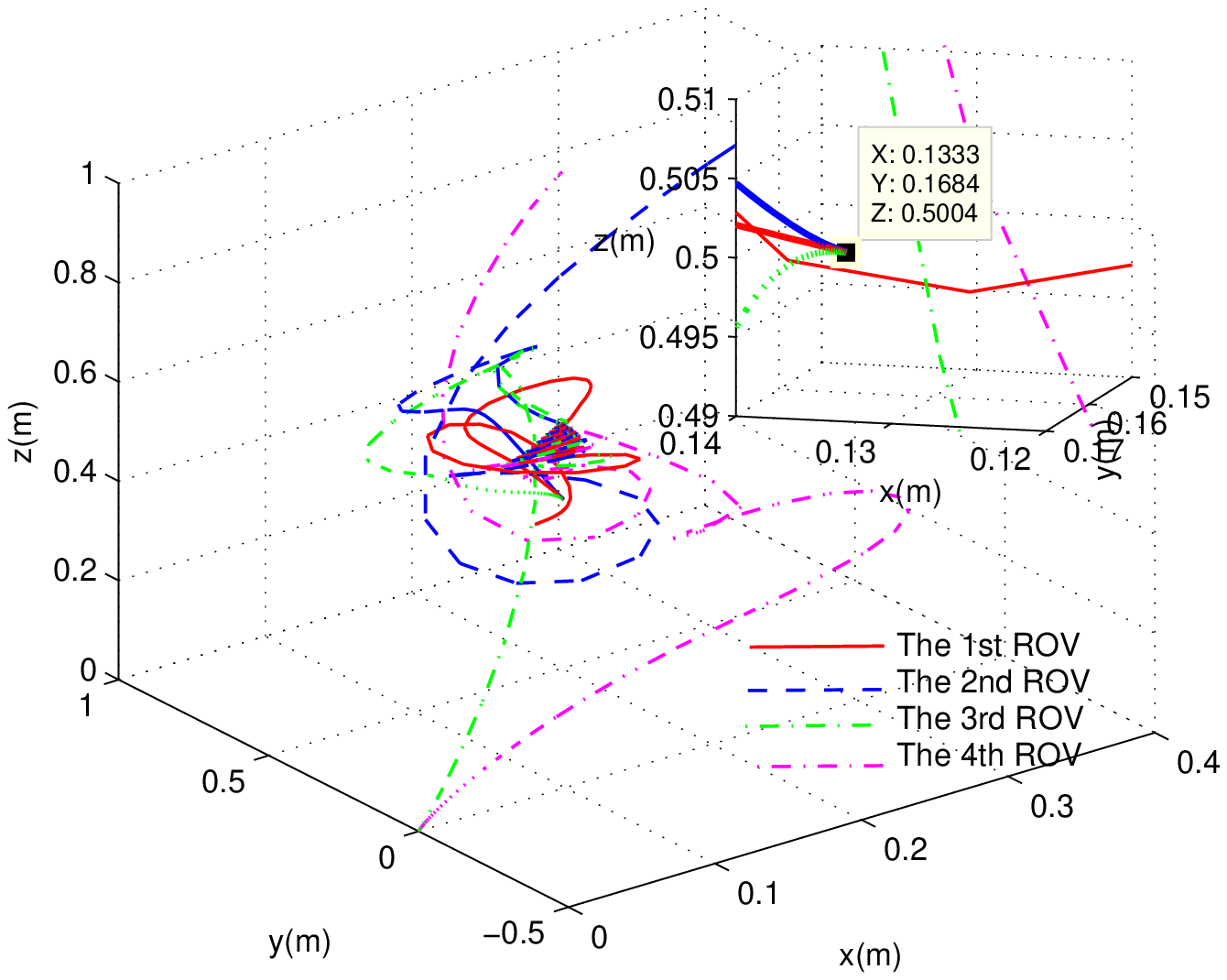}\\
  \caption{3-dimension routes of four ROVs on [0s,80s] under Case 3.}
\end{figure}

The simulation results are shown in Figs. 3-5. Fig. 3 describes the state responses of $\boldsymbol\eta^{(j)}(t)$ and $\boldsymbol\nu^{(j)}(t)$ of these four ROVs, $j=1,\cdots,4$. It can be seen that these four ROVs are arriving at the consensus at the optimal solution of (47) before the attack occurs. The ADI mechanism based on the ARR $\mho^{(j,r)}(t)$ formulated by $e_r^{(j)}$ and $e_{r,H}^{(j)}$ is shown in Fig. 4, which implies that for $j=1,2,3$, the ARR $\mho^{(j,r)}(t)$ generated by $\mathcal{M}^{(j)}$ is still satisfied even after the cyber attack occurs, while $\mho^{(4,y)}(t)$ is immediately violated (about at $t=31.5$s), thus indicating the cyber attack occurs in the 4th ROV. After detecting the cyber attack, the module $\mathcal{O}^{(j)}$ will send the control command $\Im^{(j)}=(\boldsymbol\eta_s,0,0)$ to the 4th ROV based on the secure decision-making (45). Then the ROV stops sending the transmission data $\boldsymbol\eta^{(4)}$, and converges to the secure position $\boldsymbol\eta_s=0$, while the remaining three ROVs achieve the consensus at the optimal solution of
\begin{equation}
\min_{\boldsymbol\eta^{(j)}} \sum_{j=1}^3\|\boldsymbol\eta^{(j)}-\boldsymbol\eta^{(j)}_0\|^2,~\mathrm{s.t.}~\boldsymbol\eta^{(1)}=\boldsymbol\eta^{(2)}=\boldsymbol\eta^{(3)}
\end{equation}
These have been illustrated in Fig. 3. The control inputs of the overall procedure are plotted in Fig. 5.

To further visualize the motion coordination of the ROV formation, also for comparison, the routes of the formulation of ROVs under the following three cases are presented in Figs. 7-9, respectively, where the Simulink in MATLAB running time is set as [0s,80s]:

{\bf Case 1:} Apply the basic/secure version of DOC under attack-free environment;

{\bf Case 2:} Apply the basic version of DOC (without the ADI mechanism and secure countermeasure, i.e., set $\Im^{(j)}=(y_r^{(j)},$ $\nabla g^{(j)}(y_r^{(j)}),\tilde v^{\mathbf{N}_j})$ and $F^{(j)}=0$ all the time) under adversarial environment;

{\bf Case 3:} Apply the secure version  of DOC under adversarial environment.

In Fig. 6, clearly all the ROVs rendezvous at the target position $(0.15,0.25,0.625)$ (the optimal solution of problem (47)) under the healthy condition. Fig. 7 shows that under the case when the 4th ROV is attacked, all the ROVs follow the wrong control commands and move along with wrong (insecure) routes due to the attack propagation through the exchange of information between neighboring subsystems (The Simulink reports ``ERROR'' at $t=34.45$s and terminates). However, by applying the secure DOC scheme, the ROVs 1, 2 and 3 rendezvous at the target position $(0.1333,0.1678,$ $0.5003)$ (the optimal solution of problem (48)), and the 4th ROV reach the preset secure position $(0,0,0)$, which has been illustrated in Fig. 8.

\section{Conclusions}
This paper presented a secure DOC method for a class of uncertain nonlinear CPSs. First, a basic DOC under the healthy conditions was proposed. By interacting and coordinating between cyber dynamics and physical dynamics, the consensus and optimality were guaranteed. In the presence of multiple cyber attacks, we proposed a distributed ADI approach to identify the locally occurring attacks from multiple propagating attacks. It is shown that any attack signal cannot bypass the designed ADI approach to destroy the convergence of the DOC algorithm. Finally, a secure countermeasure strategy against cyber attacks was described, which guarantees that all healthy physical subsystems complete the DOC objective, while the attacked physical subsystems converge to a given secure state.

\section*{Appendix I}

{\bf Proof of Theorem 1.} First, we give the convergence analysis on the cyber dynamics (15) and physical dynamics (1) based on the Lyapunov method, respectively.

{\bf Cyber dynamics:} Let $y^*=1_N\otimes y^\star$ be a solution of (4). By Lemma 1-(ii), there exists $v^*\in \mathbb{R}^{Nm}$ such that $\nabla g(y_r^*)+Lv^*+(1+\eta)Ly_r^*=0$ holds, and $(y^*,v^*)$ is the saddle of $G$.
Consider the Lyapunov function of the cyber dynamics
$$
V_c=\frac{1}{2}(\|y_r-y^*\|^2+\|v-v^*\|^2)
$$

Note that $z_1=y-y_r$ under the healthy conditions. Then (15) becomes
 \begin{equation}
  \begin{aligned}
\dot{y}_r=&-\nabla g(y_r)-Lv-(1+\eta)Ly_r-(1+\eta)Lz_1\\
\dot v=&Ly_r+Lz_1
\end{aligned}
\end{equation}
The time derivative of $V_c$ along with (49) is
\begin{align}
\dot V_c=\nonumber&(y_r-y^*)^T[-\nabla g(y_r)-Lv-(1+\eta)Ly_r]\\
\nonumber&+(v-v^*)^TLy_r-(1+\eta)z_1^TLy_r+(v-v^*)^TLz_1\\
\overset{(a)}{=}\nonumber&(y^*-y_r)^T[\nabla g(y_r)+Lv+Ly_r]-y_r^TLy_r\\
\nonumber&+G(y_r,v)-G(y_r,v^*)-(1+\eta)z_1^TLy_r\\
\nonumber&+(v-v^*)^TLz_1\\
\overset{(b)}{\le}\nonumber&G(y^*,v)-G(y^*,v^*)+G(y^*,v^*)-G(y,v^*)\\
\nonumber&-\eta y_r^TLy_r-(1+\eta)z_1^TLy_r+(v-v^*)^TLz_1\\
\overset{(c)}{\le}&-\eta y_r^TLy_r-(1+\eta)z_1^TLy_r+v^TLz_1-z_1^T\pi
\end{align}
where the equalities: $(a)$ follows from $Ly^*=0$ and the linearity of $G$ in its second argument; $(b)$ follows from the convexity of $G$ in the first argument; $(c)$ follows from the fact that $(y^*,v^*)$ is the saddle point of $G$. Note that the mismatching terms $v^TLz_1$ and $-z_1^T\pi$ will be compensated by the following physical dynamics.

 {\bf Physical system:} The convergence analysis is discussed based on backstepping procedure. Rewrite (1) into a compact form
\begin{equation}
\mathcal{P}:~\left\{\begin{aligned}
&\dot{x}_i=x_{i+1}+\varphi_i(\bar x_i)\theta,~i=1,\cdots,n-1\\
&\dot{x}_n=Bu+\varphi_n(x)\theta,\\
&y=x_1
\end{aligned}\right.
\end{equation}
where $\varphi_i(\bar x_i)=\mathrm{diag}\{\varphi_i^{(1)}(\bar x_i^{(1)}),\cdots,\varphi_i^{(N)}(\bar x_i^{(N)})\}$ and $\theta=\mathrm{vec}(\theta_1,\cdots,\theta_N)$.

The error dynamics can be expressed as
\begin{equation}
\left\{\begin{aligned}
&\dot{z}_1=\alpha_1+\varphi_1(\bar x_1)\theta-\dot y_r+z_2\\
&\dot{z}_i=\alpha_i+\varphi_i(\bar x_i)\theta-\dot\alpha_{i-1}+z_{i+1}\\
&\dot{z}_n=Bu+\varphi_n(x)\theta-\dot\alpha_{n-1}
\end{aligned}\right.
\end{equation}
which can be spitted into inner-loop and outer-loop subsystems:
\begin{equation}
\left\{\begin{aligned}
\dot{z}_{1,I}=&\alpha_{1,I}+\psi_1(\bar x_1)\lambda+z_2\\
\dot{z}_{i,I}=&\alpha_{i,I}+\psi_i(\bar x_i)\lambda-\dot\alpha_{i-1,I}+z_{i+1}-\mu z_i\\
\dot{z}_{n,I}=&Bu_I+\psi_n(x)\lambda-\dot\alpha_{n-1,I}-\mu z_n\\
\end{aligned}\right.
\end{equation}
and
\begin{equation}
\left\{\begin{aligned}
&\dot{z}_{1,O}=\alpha_{1,O}-\dot y_r\\
&\dot{z}_{i,O}=\alpha_{i,O}-\dot\alpha_{i-1,O}\\
&\dot{z}_{n,O}=Bu_O-\dot\alpha_{n-1,O}
\end{aligned}\right.
\end{equation}
where $z_i=z_{i,I}+z_{i,O}$ for $i=1,\cdots,n$.

Next, we provide the Lyapunov analysis of the physical dynamics by considering
$$
V_p=\frac{1}{2}\left(\sum_{i=1}^n\|z_i\|^2+\tilde\lambda^T\Gamma^{-1}\tilde\lambda+\tilde\rho^T\Gamma_0^{-1}\tilde\rho
+\tilde\pi^T\Gamma_1^{-1}\tilde\pi\right)\\
$$
where $\tilde\lambda=\lambda-\hat\lambda$, $\tilde\rho=\rho-\hat\rho$, $\tilde\pi=\pi-\hat\pi$.

The derivative of $V_p$ can be computed as
$$\begin{aligned}
\dot V_p=&\sum_{i=1}^nz_i\dot z_i-\tilde\lambda^T\Gamma^{-1}\dot{\hat\lambda}-\tilde\rho^T\Gamma_0^{-1}\dot{\hat\rho}
-\tilde\pi^T\Gamma_1^{-1}\dot{\hat\pi}\\
=&\dot V_I+\dot V_O
\end{aligned}
$$
where $\dot V_I=\sum_{i=1}^nz_i\dot z_{i,I}-\tilde\lambda^T\Gamma^{-1}\dot{\hat\lambda}-\tilde\rho^T\Gamma_0^{-1}\dot{\hat\rho}
-\tilde\pi^T\Gamma_1^{-1}\dot{\hat\pi}$ and $\dot V_O=\sum_{i=1}^nz_i\dot z_{i,O}$ represent the inner-loop and outer-loop Lyapunov derivatives, respectively.

Consider the inner-loop error dynamics (53) with controls (16)-(18) and adaptive laws (22)-(24). Following the traditional backstepping procedure \cite{MK1995}, along with (53), we can obtain
\begin{align}
\dot V_I\le\nonumber&-\sum_{i=1}^nz_i^TC_iz_i-\mu\sum_{i=2}^n\|z_i\|^2\\
&-\rho\|z_1\|^2+z_1^T\pi-z_1^TS\left(\frac{z_1}{\delta}\right).
\end{align}

Now we consider the outer-loop error dynamics (54) with controls (19)-(21).

{\it Step 1.} In view of (54) and (15), we have
\begin{align}
\dot{z}_{1,O}=\alpha_{1,O}+\nabla g(y_r)+Lv+(1+\eta)Ly
\end{align}
To stabilize (56), consider the Lyapunov derivative $\dot V_{1,O}=z_1\dot z_{1,O}$. Then using the virtual control (19), we have
\begin{align}
\dot V_{1,O}=\nonumber&z_1^T[\alpha_{1,O}+\nabla g(y_r)+Lv+(1+\eta)Ly]\\
=\nonumber&-z_1^TLv+(1+\eta)z_1^TLy\\
=\nonumber&-z_1^TLv+(1+\eta)z_1^TL(y_r+z_1)\\
\le&-z_1^TLv+(1+\eta)z_1^TLy_r+(1+\eta)\|L\|\|z_1\|^2
\end{align}
{\it Step $i (2\le i\le n)$.} Note that the arguments of the function $\alpha_{i-1,O}$ involve $y_r$ and $v$. From (54) and (15), we have
\begin{align}
\dot{z}_{i,O}\negthickspace=\nonumber&\alpha_{i,O}+\frac{\partial\alpha_{i-1,O}}{\partial y_r}\left[\nabla g(y_r)+Lv+(1+\eta)Ly\right]\\
\nonumber&-\frac{\partial\alpha_{i-2,O}}{\partial y_r}L^2y\\
=&\left[(1+\eta)\frac{\partial\alpha_{i-1,O}}{\partial y_r}L-\frac{\partial\alpha_{i-2,O}}{\partial y_r}L^2\right](y_r+z_1)
\end{align}
By using the triangular inequality, one has
$$
\begin{aligned}
(1+\eta)z_i^T\frac{\partial\alpha_{i-1,O}}{\partial y_r}Ly_r\le \nonumber&\frac{1}{4}(1+\eta)^2\|L\|z_i^T\left(\frac{\partial\alpha_{i-1,O}}{\partial y_r}\right)^2z_i\\
&+y_rLy_r\\
(1+\eta)z_i^T\frac{\partial\alpha_{i-1,O}}{\partial y_r}Lz_1\le \nonumber&\frac{1}{4}(1+\eta)^2\|L\|z_i^T\left(\frac{\partial\alpha_{i-1,O}}{\partial y_r}\right)^2z_i\\
&+\|L\|\|z_1\|^2\\
-\frac{\partial\alpha_{i-2,O}}{\partial y_r}L^2y_r\le \frac{1}{4}\|L\|^3\nonumber&z_i^T\left(\frac{\partial\alpha_{i-2,O}}{\partial y_r}\right)^2z_i+y_rLy_r\\
-\frac{\partial\alpha_{i-2,O}}{\partial y_r}L^2z_1\le \frac{1}{4}\|L\|^3\nonumber&z_i^T\left(\frac{\partial\alpha_{i-2,O}}{\partial y_r}\right)^2z_i+\|L\|\|z_1\|^2\\
\end{aligned}
$$
Also, on the compact set $\{V(t)\le V(0)\}$, there exists a positive constant such that $\|\partial\alpha_{i,O}/\partial y_r\|\le\Pi$ for all $i=1,\cdots,n-1$.
Based on these facts, the Lyapunov derivative $\dot V_{i,O}=z_i\dot z_{i,O}$ along with (58) can be expressed as
\begin{align}
\dot V_{i,O}=\nonumber&z_i^T\left[(1+\eta)\frac{\partial\alpha_{i-1,O}}{\partial y_r}L-\frac{\partial\alpha_{i-2,O}}{\partial y_r}L^2\right](y_r+z_1)\\
\le&\mu\|z_i\|^2+2\|L\|\|z_1\|^2+2y_r^TLy_r
\end{align}
Combining (57) and (59), the outer-loop Lyapunov derivative satisfies
\begin{align}
\dot V_O\le\nonumber&-z_1^TLv+(1+\eta)z_1^TLy_r+\rho\|z_1\|^2+2(n-1)y_r^TLy_r\\
&+\mu\sum_{i=2}^n\|z_i\|^2
\end{align}
Finally, construct the Lyapunov function $V=V_c+V_p$ for the overall CPS. Taking (50), (55) and (60) into account, its time derivative satisfies
\begin{align}
\dot V\le\nonumber&-(\eta-2(n-1))y_r^TLy_r-\sum_{i=1}^nz_i^TC_iz_i\\
\nonumber&-\sum_{j=1}^Nz_1^{(j)T}S\left(\frac{z_1^{(j)}}{\delta^{(j)}}\right)\\
\le&-(\eta-2(n-1))y_r^TLy_r-\sum_{i=1}^nz_i^TC_iz_i
\end{align}
where the fact $z_1^{(j)T}S(z_1^{(j)}/\delta^{(j)})\ge0$ is used.

Choose $\eta>2(n-1)$. Then $\dot V\le 0$. Thus, $\{V(t)\le V(0)\}$ is an invariant set. It implies that $z(t)$, $y_r(t)$, $v(t)$, $\hat\lambda(t)$, $\hat\rho(t)$, $\hat\pi(t)$ and $z_1^{(j)T}S(z_1^{(j)}/\delta^{(j)})$ are bounded. Then $y(t)=z_1(t)+y_r(t)$ is bounded. Along with the backstepping procedure, $\alpha_i(t)$, $u_i(t)$ and $x_i(t)$ are also bounded. Noting $\dot y_r(t),\dot v(t)\in L_\infty$ and $y_r^T(t)Ly_r(t),z_i(t)\in L_2$. According to Barbalat's
Lemma, $\lim_{t\to\infty}y_r^T(t)Ly_r(t)=0$ and $\lim_{t\to\infty}z_i(t)=0$. Finally, following the proof of [\cite{BG2014}. Theorem 4.1], one obtains that $\lim_{t\to\infty}y_r(t)=y^*$. Thus, we can conclude that $\lim_{t\to\infty}[y(t)-y^*]=\lim_{t\to\infty}[y(t)-y_r(t)+y_r(t)-y^*]=0$. $\hfill{}\blacksquare$

\section*{Appendix II}

{\bf Proof of Lemma 2.} To analyze the stability of (30), we first construct an auxiliary system
\begin{align}
\dot{\tilde e}_{r,H}^{(j)}=-\eta^{(j)}({\tilde e}_{r,H}^{(j)}+z_1^{(j)}),~{\tilde e}_{r,H}^{(j)}(0)=e_{r,H}^{(j)}(0)
\end{align}
By directly computing (62), we obtain that
$$
\|\tilde e_{r,H}^{(j)}(t)\|\le e^{-\eta^{(j)}t}e_{r,H}^{(j)}(0)+\Psi(\eta^{(j)},z_1^{(j)}(t),0,t).
$$

Also, along with (62), the time derivative of the Lyapunov function $U^{(j)}({\tilde e}_{r,H}^{(j)})=\|{\tilde e}_{r,H}^{(j)}\|^2/2$ is
\begin{align}
\dot U^{(j)}({\tilde e}_{r,H}^{(j)})=&-\eta^{(j)}{\tilde e}_{r,H}^{(j)T}({\tilde e}_{r,H}^{(j)}+z_1^{(j)})
\end{align}


On the other hand, the time derivative of $U^{(j)}(e_{r,H}^{(j)})$ along with (30) can be expressed as
\begin{align}
\dot U^{(j)}(e_{r,H}^{(j)})=\nonumber&-e_{r,H}^{(j)T}[\nabla g^{(j)}(y_r^{(j)})-\nabla g^{(j)}(\hat y_r^{(j)})]\\
\nonumber&-\eta^{(j)}e_{r,H}^{(j)T}(e_{r,H}^{(j)}+z_1^{(j)})\\
\le&-\eta^{(j)}e_{r,H}^{(j)T}(e_{r,H}^{(j)}+z_1^{(j)})
\end{align}
where the inequality follows from the convexity of $g^{(j)}$.

Using the comparison principle \cite{HK2002}, under the same initial condition ${\tilde e}_{r,H}^{(j)}(0)=e_{r,H}^{(j)}(0)$, (63) and (64) imply $U^{(j)}(e_{r,H}^{(j)})\le U^{(j)}({\tilde e}_{r,H}^{(j)})$, or equivalently, $\|e_{r,H}^{(j)}(t)\|\le\|{\tilde e}_{r,H}^{(j)}(t)\|\le e^{-\eta^{(j)}t}e_{r,H}^{(j)}(0)+\Psi(\eta^{(j)},z_1^{(j)}(t),t_0,t)$.

Directly solving (31) and applying the triangular inequality, $\|e_{v,H}^{(j)}(t)\|$ can be bounded by (33).

\section*{Appendix III}

{\bf Proof of Lemma 3.} Under the healthy conditions, integrating both sides of Eq. (61) yields
$$
\begin{aligned}
V(t)-V(0)\le& -(\eta-2n+2)\int_{\tau=0}^t y_r^T(\tau)Ly_r(\tau)d\tau\\
&-\sum_{i=1}^n\int_{\tau=0}^t z_i^T(\tau)C_i z_i(\tau)d\tau
\end{aligned}
$$
which implies that
$$
\begin{aligned}
\frac{1}{2}\| z_1^{(j)}(t)\|^2+c_1^{(j)}\int_{\tau=0}^t\| z_1^{(j)}(\tau)\|^2d\tau\le V(0).
\end{aligned}
$$
Given Assumption 3 and definition of $\bar\Omega$, we have
$$
\frac{1}{2c_1^{(j)}}\| z_1^{(j)}(t)\|^2+\int_{\tau=0}^t\| z_1^{(j)}(\tau)\|^2d\tau\le\bar\Omega/c_1^{(j)}
$$
In addition, from Theorem 1 we know $z_1^{(j)T}S(z_1^{(j)}/\delta^{(j)})$ is bounded, which yields $|z_{1,s}^{(j)}(t)|<\delta^{(j)}(t)$ for any $t\ge 0$. To prove the result we suppose, for contradiction, there exists $s\in\{1,\cdots,m\}$ such that $|z_{1,s}^{(j)}(t)|\ge\delta^{(j)}(t)>k_b^{(j)}$. According to the form of $S$ and the prescribed performance technique \cite{WW2010}, $\|S(z_1^{(j)}/\delta^{(j)})\|$ converges to $+\infty$. Then $z_1^{(j)T}S(z_1^{(j)}/\delta^{(j)})\ge k_b^{(j)}\|S(z_1^{(j)}/\delta^{(j)})\|\to+\infty$ as $t\to\infty$, a contradiction, which in turn implies $\|z_1^{(j)}(t)\|<\sqrt{m}\delta^{(j)}(t)$.

%
%


\end{document}